\documentclass[twocolumn]{aastex631}
\usepackage{natbib}
\received{March 14, 2023}
\revised{May 15, 2023}
\accepted{May 26, 2023}

\shorttitle{Narrow-line Seyfert 1 galaxy with VERA polarimetric observations}
\shortauthors{Takamura et al.}
\graphicspath{{./}{figures/}}

\begin{document}

\title{Probing the Heart of Active Narrow-line Seyfert 1 Galaxies with VERA Wideband Polarimetry}

\correspondingauthor{Mieko Takamura}
\email{mieko.takamura@grad.nao.ac.jp}

\author[0000-0002-5556-1855]{Mieko Takamura}
\affiliation{Department of Astronomy, Graduate School of Science, The University of Tokyo, 7-3-1 Hongo, Bunkyo, Tokyo 113-0033, Japan}
\affiliation{Mizusawa VLBI Observatory, National Astronomical Observatory of Japan, 2-12 Hoshigaoka, Mizusawa,
Oshu 023-0861, Japan}

\author[0000-0001-6906-772X]{Kazuhiro Hada}
\affiliation{Mizusawa VLBI Observatory, National Astronomical Observatory of Japan, 2-12 Hoshigaoka, Mizusawa,
Oshu 023-0861, Japan}
\affiliation{Department of Astronomical Science, The Graduate University for Advanced Studies (SOKENDAI),
2-21-1 Osawa, Mitaka 181-8588, Japan}

\author[0000-0003-4058-9000]{Mareki Honma}
\affiliation{Mizusawa VLBI Observatory, National Astronomical Observatory of Japan, 2-12 Hoshigaoka, Mizusawa,
Oshu 023-0861, Japan}

\author[0000-0003-4046-2923]{Tomoaki Oyama}
\affiliation{Mizusawa VLBI Observatory, National Astronomical Observatory of Japan, 2-12 Hoshigaoka, Mizusawa,
Oshu 023-0861, Japan}

\author{Aya Yamauchi}
\affiliation{Mizusawa VLBI Observatory, National Astronomical Observatory of Japan, 2-12 Hoshigaoka, Mizusawa,
Oshu 023-0861, Japan}

\author{Syunsaku Suzuki}
\affiliation{National Astronomical Observatory of Japan, 2-21-1 Osawa, Mitaka 181-8588, Japan}

\author[0000-0002-9043-6048]{Yoshiaki Hagiwara}
\affiliation{Natural Science Laboratory, Toyo University, 5-28-20 Hakusan, Bunkyo-ku, Tokyo 112-8606, Japan}
\affiliation{ASTRON and Joint Institute VLBI for ERIC (JIVE), Oude Hoogeveensedijk 4, 7991 PD Dwingeloo, The Netherlands}

\author[0000-0003-4470-7094]{Monica Orienti}
\affiliation{INAF Istituto di Radioastronomia, via Gobetti 101, I-40129 Bologna, Italy}

\author[0000-0001-7618-7527]{Filippo D'Ammando}
\affiliation{INAF Istituto di Radioastronomia, via Gobetti 101, I-40129 Bologna, Italy}

\author[0000-0001-6558-9053]{Jongho Park}
\affiliation{Korea Astronomy and Space Science Institute, Daedeok-daero 776, Yuseong-gu, Daejeon 34055, Republic of Korea}

\author[0000-0001-9799-765X]{Minchul Kam}
\affiliation{Department of Physics and Astronomy, Seoul National University, Gwanak-gu, Seoul 08826, Republic of Korea}

\author[0000-0003-4384-9568]{Akihiro Doi}
\affiliation{Institute of Space and Astronautical Science, Japan Aerospace Exploration Agency, 3-1-1 Yoshinodai, Chuo, Sagamihara 252-5210, Japan}
\affiliation{Department of Space and Astronautical Science, SOKENDAI, 3-1-1 Yoshinodai, Chuou-ku, Sagamihara, Kanagawa 252-5210, Japan}

\begin{abstract}
We explored the parsec-scale nuclear regions of a sample of radio-loud narrow-line Seyfert 1 galaxies (NLSy1s) using the VLBI Exploration of Radio Astronomy (VERA) wideband (at a recording rate of $16\,\mathrm{Gbps}$) polarimetry at 22 and 43\,GHz. Our targets include 1H\,0323+342, SBS\,0846+513, PMN\,J0948+0022, 1219+044, PKS\,1502+036 and TXS\,2116-077, which are all known to exhibit $\gamma$-ray emission indicative of possessing highly beamed jets similar to blazars. For the first time, we unambiguously detected Faraday rotation toward the parsec-scale radio core of NLSy1s, with a median observed core rotation measure (RM) of $2.7\times 10^3\,{\rm rad\,m^{-2}}$ (or $6.3\times 10^3\,{\rm rad\,m^{-2}}$ for redshift-corrected). This level of RM magnitude is significantly larger than those seen in the core of BL Lac objects (BLOs; a dominant subclass of blazars), suggesting that the nuclear environment of NLSy1s is more gas-rich than that in BLOs. Interestingly, the observed parsec-scale polarimetric properties of NLSy1s (low core fractional polarization, large core RM and jet--EVPA misalignment) are rather similar to those of flat-spectrum radio quasars (FSRQs). Our results are in accordance with the scenario that NLSy1s are in an early stage of AGN evolution with their central black hole masses being smaller than those of more evolved FSRQs.
\end{abstract}



\section{Introduction} \label{sec:intro}
Narrow line Seyfert 1 galaxies (NLSy1s) are a subclass of active galactic nuclei (AGNs). This class is defined by its optical properties, i.e., narrow permitted line FWHM (H$\beta < 2000\, \mathrm{km\,s^{-1}}$), $[\mathrm{OIII}] / \mathrm{H} \beta<3$, and a strong bump due to $\mathrm{FeII}$ emission lines \citep{Osterbrock1985}. They also exhibit strong X-ray variability with substantial excess in soft X-rays and relatively high luminosity \citep[e.g,][]{Boller1996}. 
These characteristics suggest that NLSy1s have smaller masses of the central black hole ($10^6$--$10^8$$M_{\odot}$) and higher accretion rates (close to or up to the Eddington limit) than those of other classes of AGNs such as quasars, BL Lac objects (BLOs) and radio galaxies, rendering NLSy1s a unique laboratory to study the physics of rapidly-evolving supermassive black holes (SMBH). 
 
 Although a majority of NLSy1s are radio-quiet, $\sim$7\% of them are radio-bright objects \citep{Zhou2006, Yuan2008, Rakshit2017}. 
 Previous studies in radio bands have reported that radio-loud NLSy1s (RL-NLSy1s) tend to have properties such as steep or flat radio spectra \citep{Yuan2008, Doi_2011,Foschini2015, Gu2015, Shao2023}, significant variability~\citep{Angelakis2015, D'Ammando2013}, high-brightness-temperature compact radio core on parsec scales \citep{Yuan2008, Wajima2014, Gu2015, Doi2016},  and ejection of superluminal jet features \citep{D'Ammando2013, Lister2016, Doi2018,Lister2019}. These characteristics are all reminiscent of blazars, i.e., a class of radio-loud AGNs with their relativistic jets pointing to our light-of-sight. Moreover, recent \textit{Fermi}-LAT observations have identified $\gamma$-ray emission from a dozen of the brightest RL-NLSy1s \citep{Abdo2009a, Abdo2009b,D'Ammando2019}. The discovery of $\gamma$-ray emission from NLSy1s indicates the presence of a third class of powerful jetted AGN along with blazars and radio galaxies. This poses a challenge to our current understanding of AGN jet formation, as NLSy1s are thought to be hosted by late-type (spiral) galaxies, which are not expected to produce high-power jets. 
 
 In order to better understand the accretion and ejection as well as their connection to the nuclear environment around SMBH, high-resolution very-long-baseline interferometry (VLBI) with full polarimetric capability is particularly useful. Polarimetric VLBI enable the measurement of the degree of polarization and electric-vector-polarization angles (EVPA) of the synchrotron emitting plasma on parsec scales, which are closely associated with the structure of magnetic fields near SMBH. Moreover, since polarized waves are affected by Faraday rotation when propagating through non-relativistic plasma within or external to the source~\citep[e.g,][]{Burn1996, Hovatta2012}, one can derive the rotation measure (RM) as $\chi_{\rm obs} \propto \mathrm{RM}\mathrm{\lambda^2}$ (if EVPAs are measured in multiple bands), which serves as a probe of physical conditions (strength of magnetic field, electron density and spatial extent) of the Faraday screen surrounding the emission region. In fact, previous systematic VLBI RM studies on bright radio-loud AGNs \citep{Hovatta2012, Park2018} found that flat-spectrum-radio quasars (FSRQs: a high-luminosity subclass of blazars) tend to have higher RM values than those of BLOs (a low-luminosity subclass of blazars), suggesting that FSRQs have a more gas-rich nuclear environment than BLOs. Nevertheless, polarimetric VLBI observations on (RL-)NLSy1s are still severely limited, mainly due to their weaker emissions than those from brighter AGNs. Moreover, previous VLBI observations of NLSy1s were largely limited to low ($\lesssim$15\,GHz) frequencies except for a few rare examples~\citep[e.g.,][]{Doi2016, Hada2018}, precluding the access to the closer vicinity of the central engine due to large optical depths for synchrotron emission.  
 
In this paper, we present the first systematic high-frequency (22 and 43\,GHz) and polarimetric VLBI study for a sample of RL-NLSy1s using the VLBI Exploration of Radio Astrometry (VERA), a VLBI network consisting of four 20\,m radio telescopes distributed in Japan. As summarized in \citet{Hagiwara2022}, recently, VERA has significantly upgraded its observing capability by installing dual-polarization receiving frontend and very wideband (16\,Giga\,bit\,s$^{-1}$) recording backend systems. This enables us to explore the polarimetric properties of weaker radio sources in considerable detail. 

The paper is organized as follows: In Section 2, we describe our sample. In Section 3, we describe our VERA observations and data reduction. In Section 4, we present the results (both total intensity and polarimetric ones). In Section 5, we discuss the nuclear properties of NLSy1s based on the new results, particularly in comparison with other types of AGNs. In the final section, we summarize the paper. Throughout this paper, we adopt $H_{\rm 0}=67.7{\rm \,km\,s^{-1}\,Mpc^{-1}}$, $\Omega_{\rm M} = 0.31$, and $\Omega_{\rm \Lambda} = 0.69$~\citep{Planck2020}.

\section{Sample} 
We selected the sample for this study as follows: First, out of the nine ``genuine" $\gamma$-ray detected NLSy1s registered in the Fourth Catalog of \textit{Fermi}-LAT of Sources \citep[4FGL;][]{Abdollahi2020}, we selected the sources that satisfied the following two conditions so that VERA can detect them at sufficient signal-to-noise ratios (SNRs): (1) J2000.0 declination $\geq$$-10^{\circ}$ and (2) milliarcsecond (mas)-scale Stokes $I$ values $\gtrsim$100\,mJy reported in previous VLBI literatures. This leaves 1H 0323+342, SBS 0846+513, PMN J0948+0022, PKS 1502+036 and TXS 2116-077. In addition, we also included 1219+044, a bright $\gamma$-ray emitting radio-loud AGN whose spectral type is claimed to be a possible candidate of NLSy1 in the literature \citep{Kynoch2019}. 
The whole list of our sample is summarized in Table \ref{tab:sources}.

\section{Observations and Data reductions} \label{sec:data}

\subsection{VERA wideband polarimetric observations}\label{ssec:obs}
On 2022 March 8 and 13, we made VERA polarimetric observations for a total of 11 sources in K and Q bands\footnote{Hereafter we refer to 22/43\,GHz frequency bands as K/Q bands since our actual receiving frequency range in each band exceeds 1\,GHz.}, respectively. The sources consist of 6 NLSy1 targets and 5 bright calibrators as summarized in Table\,\ref{tab:sources}. Each session was conducted for a continuous 19-hour track, where all the 11 sources were observed over a wide range of parallactic angles by alternating sources every 5--15 minutes to improve $(u,v)$-coverage and polarimetric calibration accuracy. The total on-source time of each target was typically 120--180 minutes, while that of each calibrator was 60 minutes. All the four VERA stations participated in good weather conditions throughout the sessions, with typical system temperatures of 100--200\,K in K band and 200--800\,K in Q band, respectively. Although VERA is capable of a dual-beam mode that is used for astrometric observations~\citep[e.g.,][]{Vera2020}, here we employed a single-beam mode that recorded data from one of the two receivers. 

Our VERA observations were performed with an ultra-wideband recording mode of 16\,Gbps (8\,Gbps for each left-/right-hand-circular polarization (LCP/RCP)), where a new digital backend system consisting of a high-speed analog-to-digital sampler OCTAD and a high-speed data recorder OCTADISK2 were employed (see \citealt{Oyama2016, Hagiwara2022} for details). For a 2-bit signal quantization, this enabled us to record data for a total bandwidth of 2048\,MHz per polarization. We received a frequency range of 21459--23507\,MHz for K band and 42426--44474\,MHz for Q band, respectively. The down-converted signals were then filtered to four 512\,MHz basebands per polarization. The data were correlated with a software FX-type correlator installed at the Mizusawa VLBI Observatory~\citep[{\tt softcos}/OCTACOR2;][]{Oyama2016}. Our observations are summarized in Table \ref{tab:observations}.

\subsection{Data reduction and polarimetric calibration}\label{ssec:reduction}
As VERA polarimetric data are obtained with a single-beam mode (in contrast to a dual-beam mode that is regularly used for astrometric observations), the data reduction procedures are similar to those of other existing dual-polarization VLBI arrays such as the Very-Long-Baseline Array (VLBA) and the Korean VLBI network (KVN). The initial data calibration was performed with the Astronomical Image Processing System (AIPS, \citealt{Greisen2003}) developed at the National Radio Astronomy Observatory (NRAO). First, apriori amplitude calibration was applied to each antenna using the antenna elevation-gain curve and (opacity-corrected) system temperatures. 
For phase calibration, we first corrected known phase variations due to parallactic angle effects. Next, the instrumental phase and delay offsets for each antenna/subband were derived using scans of bright calibrators. We then performed a global fringe fitting to the whole data with a solution interval of 1 minute, removing delay, rate and phase residuals in each 512\,MHz subband separately. We detected fringes for most of the scans (both targets and calibrators) at adequate SNRs. Subsequently, the cross-hand R-L phase and delay offsets and complex bandpass of each antenna/subband were calibrated by using scans brightest calibrators (3C\,84 or OJ\,287). Following the above initial calibration in AIPS, CLEAN imaging of Stokes $I$, $Q$ and $U$ on each source were performed with the Caltech Difmap package \citep{Shepherd1997}. For the calibration of antenna complex gains, self-calibration of phase/amplitude was performed using the Stokes $I$ data set. 

We derived the feed polarization leakages for each antenna (so-called ``D-terms") in AIPS using a sophisticated calibration package, GPCAL~\citep{Park2021}, which enables us to fit multiple sources simultaneously and significantly improves the determination accuracy of D-terms compared to the traditional calibration task, LPCAL~\citep[e.g.,][]{Leppanen1995}. We used PKS\,0235+164, 3C\,84 and OJ\,287 to obtain the initial coarse solutions of D-terms, and subsequently added 5 more sources and iterated the calibration until the estimations of the D-terms were well converged. For both K and Q bands, D-terms were derived for each of the four 512\,MHz subbands separately to take into account the possible frequency dependence of D-terms over such a wide passband. For more details of D-term estimation with GPCAL, see Appendix \ref{sec:appendix-dterm}. In summary, the final D-terms were derived to be $<$10\% in K band and $<$15\% in Q band for all the four VERA stations, respectively.

\begin{deluxetable}{cccc}
\tablenum{1}
\tablecaption{Targets and calibrators of our observations\label{tab:sources}}
\tablewidth{0pt}
\tablehead{
\colhead{Source} &\colhead{Class}&\colhead{$z$}&\colhead{$M_{\rm BH}(\times10^7 M_{\odot}$)}}
\decimalcolnumbers
\startdata
1H\,0323+342 &NLSy1&0.061\textsuperscript{(a)}&$2$\textsuperscript{(e)}\\
SBS\,0846+513&NLSy1& 0.584 \textsuperscript{(a)}&$3.9$\textsuperscript{(f)}\\
PMN\,J0948+0022&NLSy1& 0.585 \textsuperscript{(a)}&$3.2$\textsuperscript{(f)} \\
1219+044&NLSy1& 0.966 \textsuperscript{(c)} &$20$\textsuperscript{(g)}\\
PKS\,1502+036&NLSy1& 0.408 \textsuperscript{(a)}&$0.18$\textsuperscript{(f)}\\
TXS\,2116-077&NLSy1& 0.260 \textsuperscript{(b)} &$1.6$\textsuperscript{(f)}\\
\hline
PKS\,0235+164&BLO&0.94\textsuperscript{(d)}&$13$ \textsuperscript{(l)}\\
OJ\,287&BLO&0.31\textsuperscript{(d)}&$1800$ \textsuperscript{(k)}\\
0528+134&FSRQ&2.06\textsuperscript{(d)}&$450$ \textsuperscript{(i)}\\
3C\,273&FSRQ&0.158\textsuperscript{(d)}&$ 26$ \textsuperscript{(h)}\\
3C\,84&RG&0.0176\textsuperscript{(d)}&$90$ \textsuperscript{(j)}\\
\hline
\enddata
\tablecomments{(1) Source.  (2) NLSy1: Narrow-line Seyfert 1 galaxy; FSRQ: Flat-spectrum radio quasar; BLO: BL Lac object; RG: Radio galaxy. (3) Redshift (\textsuperscript{(a)}\citealp{D'Ammando2020}; \textsuperscript{(b)}\citealp{Romano2018}; \textsuperscript{(c)}\citealp{Kynoch2019}; \textsuperscript{(d)}NASA/IPAC Extragalactic Database (NED)) (4) Black hole mass (\textsuperscript{(e)}\citealp{Zhou2007}; \textsuperscript{(f)}\citealp{Rakshit2017}; \textsuperscript{(g)}\citealp{Kynoch2019}; \textsuperscript{(h)}\citealp{Sturm2018}; \textsuperscript{(i)}\citealp{Yan2012}; \textsuperscript{(j)}\citealp{Scharwachter2013}; \textsuperscript{(k)}\citealp{Valtonen2012};\textsuperscript{(l)}\citealp{Gupta2012})  }
\end{deluxetable}

\begin{deluxetable*}{cccc}
\tablenum{2}
\tablecaption{Summary of VERA wideband polarimetric observations\label{tab:observations}}
\tablewidth{0pt}

\tablehead{
\colhead{Project code} &\colhead{Observing date}&\colhead{Antenna}& \colhead{Frequency (MHz)}}
\decimalcolnumbers
\startdata
R22067A& 2022 Mar 8 23:55 - Mar 9 18:55  & MIZ, IRK, OGA, ISG & 21715, 22270, 22739, 23251 \\
R22072B& 2022 Mar 13 23:54 - Mar 13 18:54 & MIZ, IRK, OGA, ISG& 42682, 43194, 43706, 44218\\
\hline
\enddata
\tablecomments{(1) Project ID code of VERA observations. (2) Observing date and time in UT. (3) MIZ: Mizusawa, IRK: Iriki, OGA: Ogasawara, ISG: Ishigaki-jima. (4) Central radio frequency (in MHz) in each of the four 512\,MHz subbands.}
\end{deluxetable*}

\begin{deluxetable*}{cccccccccc}
    \tablenum{3}
    \tablecaption{Summary of each polarized image parameter}
    \label{tab:imageparameters}
    \tablewidth{0pt}
    \tablehead
    {
    \colhead{Epoch}&\colhead{Source}&\colhead{$t_{int}$}&\colhead{$I_{peak}$}&\colhead{$I_{tot}$}& \colhead{Beam size}&\colhead{$I_{rms}$}&\colhead{DR}&\colhead{$P_{peak}$}&\colhead{$\sigma_{P}$}\\
    \colhead{}&\colhead{}&\colhead{(min)}&\colhead{($\frac{\mathrm{mJy}} {\mathrm{beam}}$)}& \colhead{($\mathrm{mJy}$)}& \colhead{(mas,mas,deg)}&\colhead{($\frac{\mathrm{mJy}}{\mathrm{beam}}$)}& \colhead{}&\colhead{($\frac{\mathrm{mJy}}{\mathrm{beam}}$)}&\colhead{($\frac{\mathrm{mJy}}{\mathrm{beam}}$)}
    }
    \decimalcolnumbers
\startdata
    R22067A & 1H\,0323+342 &196 & 135 & 144 & 0.90, 1.32, $-$41.7  & 0.12  & 1108&3.3 &0.2\\
    (Kband)& SBS\,0846+513 &134 & 276 & 299 & 0.91, 1.25, $-$50.2  & 0.11  & 2437&7.5 & 0.2\\
        & PMN\,J0948+022 &100 & 495 & 506 & 0.87, 1.57, $-$34.3   & 0.12  & 4274& 3.9 & 0.3\\
        & PKS\,1502+036 &90 & 373 & 393 & 0.80 1.71, $-$34.0  & 0.40  & 938 & 7.7  & 0.3\\
        & TXS\,2116-077 &112 & 128 & 133 & 0.81, 1.67, $-$26.5  & 0.16  & 782 & 5.4  & 0.1\\
        & 1219+044 &55 & 588 & 612 & 0.87, 1.55, $-$37.0   & 0.23  & 2580 & 18 & 0.3\\
        & 0528+134 &63 & 1065 & 1077 & 0.83, 1.49, $-$36.4 & 1.1  & 968 & 24  & 0.4\\
        & 3C\,273 &50 & 5829 & 9787 & 0.79, 1.78, $-$36.6  & 9.3  & 627 & 112 & 2.2\\
        & OJ\,287 &64 & 5855 & 6307 & 0.86, 1.42, $-$40.6   & 6.1  & 960 & 520 & 2.3\\
        & PKS\,0235+164 &60 & 1470 & 1500 & 0.84, 1.53, -36.5  & 2.5  & 588 & 62 & 0.9 \\
        & 3C\,84 &60 & 5869 & 14904 & 0.85, 1.35, $-$31.0   & 46.7  & 126 & 39 & 3.1 \\
    \hline
    R22072B & 1H 0323+342 &195 & 125 & 132 & 0.44, 0.70, $-$37.6   & 0.29  & 433& 2.8  & 0.4\\
    (Qband)& SBS 0846+513 &140 & 218 & 233 & 0.50, 0.67, $-$25.2  & 0.46 & 425 & 6.3 & 0.4\\
        & PMN\,J0948+022 &104 & 361 & 367 & 0.43, 0.92, $-$36.5  & 0.43  & 836 & 5.6 & 1.1\\
        & PKS\,1502+036 &68 & 224 & 237 & 0.37, 0.99, $-$33.0 & 3.01  & 74&--&--\\
        & TXS\,2116-077 &130 & 70.3 & 88.5 & 0.53, 0.76, $-$26.8  & 2.36  & 30&--&--\\
        & 1219+044 &65 & 282 & 288 & 0.40, 0.90, $-$37.3   & 0.86  & 328& 8.9 & 1.1\\
        & 0528+134 &65 & 716 & 750 & 0.43, 0.78, $-$38.2  & 0.79  & 906& 20 & 0.5\\
        & 3C\,273 &50 & 2020 & 4102 & 0.42, 0.82, $-$35.2   & 15.9  & 127& 20 & 1.2\\
        & OJ\,287 &51 & 3464 & 4242 & 0.44, 0.73, $-$34.3  & 5.0  & 693& 406& 1.1\\
        & PKS\,0235+164 &60 & 993 & 1060 & 0.43, 0.75, $-$39.6   & 1.13  & 879& 32  & 0.6 \\
        & 3C\,84 &60 & 3542 & 6322 & 0.48, 0.65, $-$49.1   & 26.4  & 134& 20  & 1.3\\
    \hline
 \enddata
    \tablecomments{(1) Epoch. (2) Source. (3) Total integration time on each source. (4) Image peak intensity. (5) Milliarcsecond-scale integrated flux density. (6) Beam size for each source. (7) Stokes I image rms noise level. (8) Image dynamic range. (9) Peak polarized intensity. (10) Polarized image rms noise. }
\end{deluxetable*}

\section{Results}
\subsection{Total intensity and polarization images}\label{ssec:images}
\begin{figure*}[htbp]
 \begin{center}
 \hspace{-5mm}
  \includegraphics[width=1.0\linewidth]{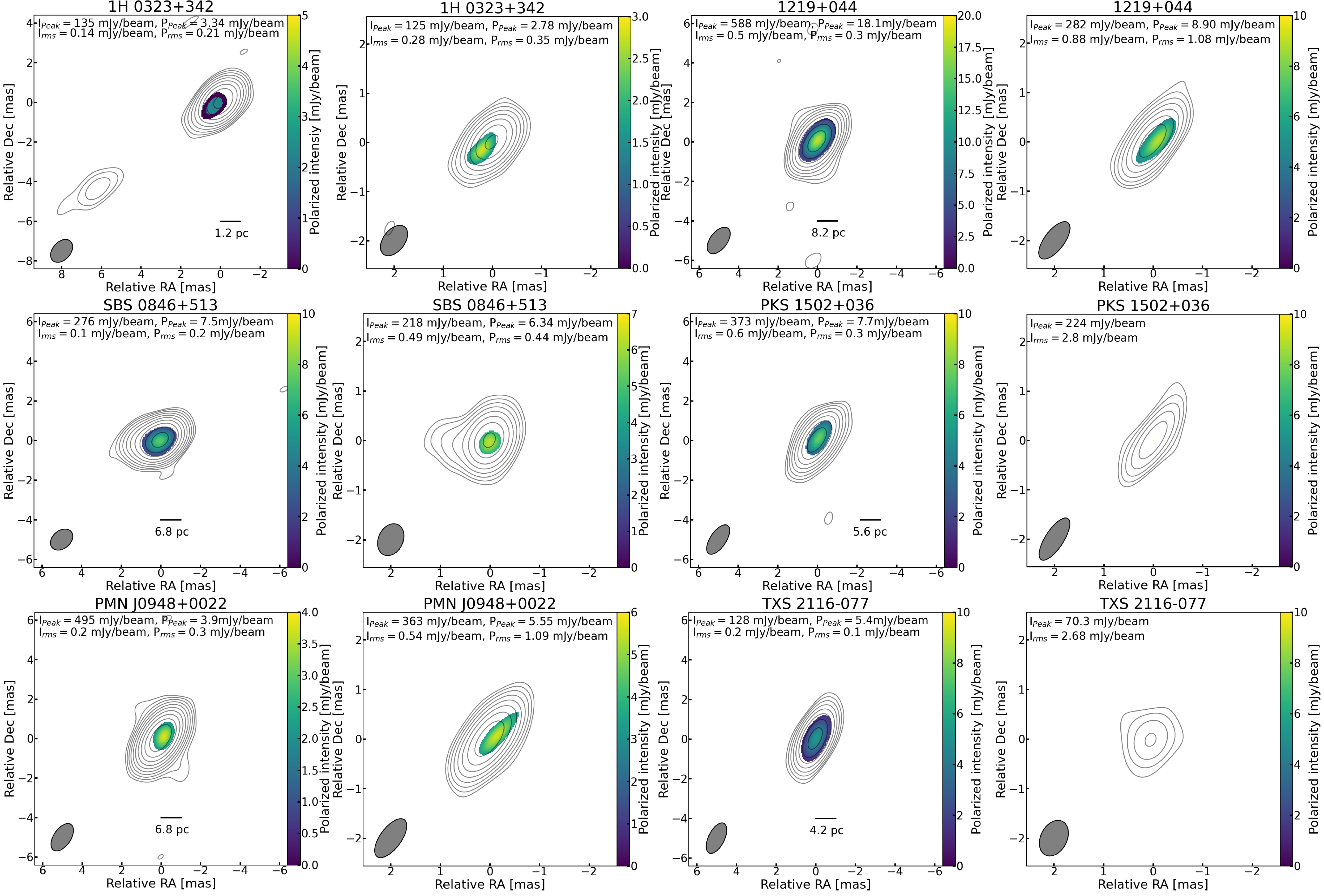}
  \caption{Total intensity and linear polarization images of NLSy1s. Contours indicate the total intensity and color scales show the polarized intensity. Panels in the first and third columns are K band images, while panels in the second and fourth columns are Q band images.}
  \label{fig:images}
 \end{center}
\end{figure*}

In Figure\,\ref{fig:images}, we show linear-polarization images of the six NLSy1 targets obtained with VERA at K and Q bands, overlaid on Stokes $I$ intensity contours. For completeness, we also show the corresponding images of our calibrators in Appendix \ref{sec:polimage}. To highlight relatively weak polarization signals, we show the images in Figure\,\ref{fig:images} integrated over the whole four 512\,MHz subbands in each K/Q band, which minimizes the image rms noise levels. The image parameters for these maps are summarized in Table\,\ref{tab:imageparameters}.

We obtain the Stokes $I$ images for all the targets at both frequencies. For the five sources, except for 1H\,0323+342 that was previously studied in detail, this is the first VLBI detection and imaging in Stokes $I$ up to K/Q bands, resolving the parsec-scale structures of these active nuclei. The parsec-scale structures of these sources appear to be largely compact, but 1H\,0323+342 and SBS\,0846+513 exhibit extended or slightly elongated morphology toward the direction of large-scale jets known in their previous lower-frequency VLBI images. 

In Table~\ref{tab:spectral}, we summarize the spectral index ($\alpha_{\rm KQ}$) and the observer-frame brightness temperature ($T_{B,\mathrm{obs}}$) measured for the core region of each source. $\alpha_{\rm KQ}$ is calculated by comparing the K/Q-bands flux densities integrated over a squared area of 1\,mas$^2$ centered on the radio core. $T_{B,\mathrm{obs}}$ is calculated at Q band (owing to the higher resolution) using the measured peak intensity of the radio core with a nominal beam size; hence, the values derived in this formula can be regarded as a lower limit if the radio core is more compact than the beam size. One can see that all the six NLSy1 targets exhibit high brightness temperature ($T_{B,\mathrm{obs}}\gtrsim10^8$\,K) radio cores with flat-to-steep spectra in these frequency bands, similar to classical blazars \citep{Yuan2008, Doi_2011, Foschini2015, Gu2015, Shao2023}.

As for polarization, owing to the wideband recording capability of VERA, we successfully detect weak polarization features in most of our NLSy1 targets at high SNRs. In K band, polarization signals are robustly detected in all six targets at peak-to-noise ratios of 13--60. In Q band, the observing conditions are overall more challenging. We detect significant polarization fluxes in four out of the six sources, whereas in the other two targets (TXS\,2116-077 and PKS\,1502+036) we fail to see polarized fluxes. We had instrumental issues during the scans of these two sources. This significantly degraded our Q-band image noise levels in these two sources (as seen in Table\,\ref{tab:imageparameters}), causing poorer sensitivity to polarization. For all the targets, the location of peak polarization signals appears to be largely coincident with the total-intensity core (see Section\,\ref{ssec:individual} for notes on individual sources). 

We also estimate the mas-scale fractional polarization degree ($m = \frac{\sqrt{Q^2+U^2}}{I}$) by comparing the polarized and total-intensity fluxes in a common area. For K and Q band images, the measured areas were set to be 2$\times$2\,mas$^2$ and 1$\times$1\,mas$^2$ centered on the core, respectively. As summarized in Table\,\ref{tab:rm}, we find that the mas-scale fractional polarization degrees are rather modest (no larger than 5\%) for all the six NLSy1 targets. A tendency of low core fractional polarization in NLSy1s was also reported in a previous study based on Monitoring Of Jets in Active galactic nuclei with VLBA Experiments (MOJAVE) at 15\,GHz~\citep{Hodge2018}.

\begin{deluxetable}{ccc}
\tablenum{4}
\tablecaption{Spectral index and observed brightness temperature of the radio core \label{tab:spectral}}
\tablewidth{0pt}
\tablehead{
\colhead{Source} &\colhead{$\alpha_{\rm KQ}$}&\colhead{$T_{B, \mathrm{obs}} (\mathrm{K}$)}}
\decimalcolnumbers
\startdata
1H\,0323+342& $-$0.07 & $2.7\times10^8$\\
SBS\,0846+513&$-$0.34 & $4.3\times10^8$\\
PMN\,J0948+0022&$-$0.46 & $6.1\times10^8$\\
PKS\,1502+036&$-$0.77&$4.0\times10^8$\\
TXS\,2116-077&$-$0.61&$1.1\times10^8$\\
1219+044&$-$1.13 &$5.2\times10^8$\\
\hline
0528+134&$-$0.55&$1.4\times10^9$ \\
3C273&$-$1.22 &$9.5\times10^{9}$\\
OJ287&$-$0.60&$7.1\times10^9$\\
PKS\,0235+164&$-$0.54&$2.2\times10^9$\\
3C84&$-$0.53&$7.5\times10^{9}$\\
\hline
\enddata
\tablecomments{(1) Source name. (2) Spectral index between K and Q bands. (3) Observed brightness temperature at 43\,GHz. }
\end{deluxetable}

\subsection{Inband RM within K-band}\label{sec:inbandrm}

\begin{figure*}[htbp]
 \begin{center}
  \includegraphics[width=1.0\linewidth]{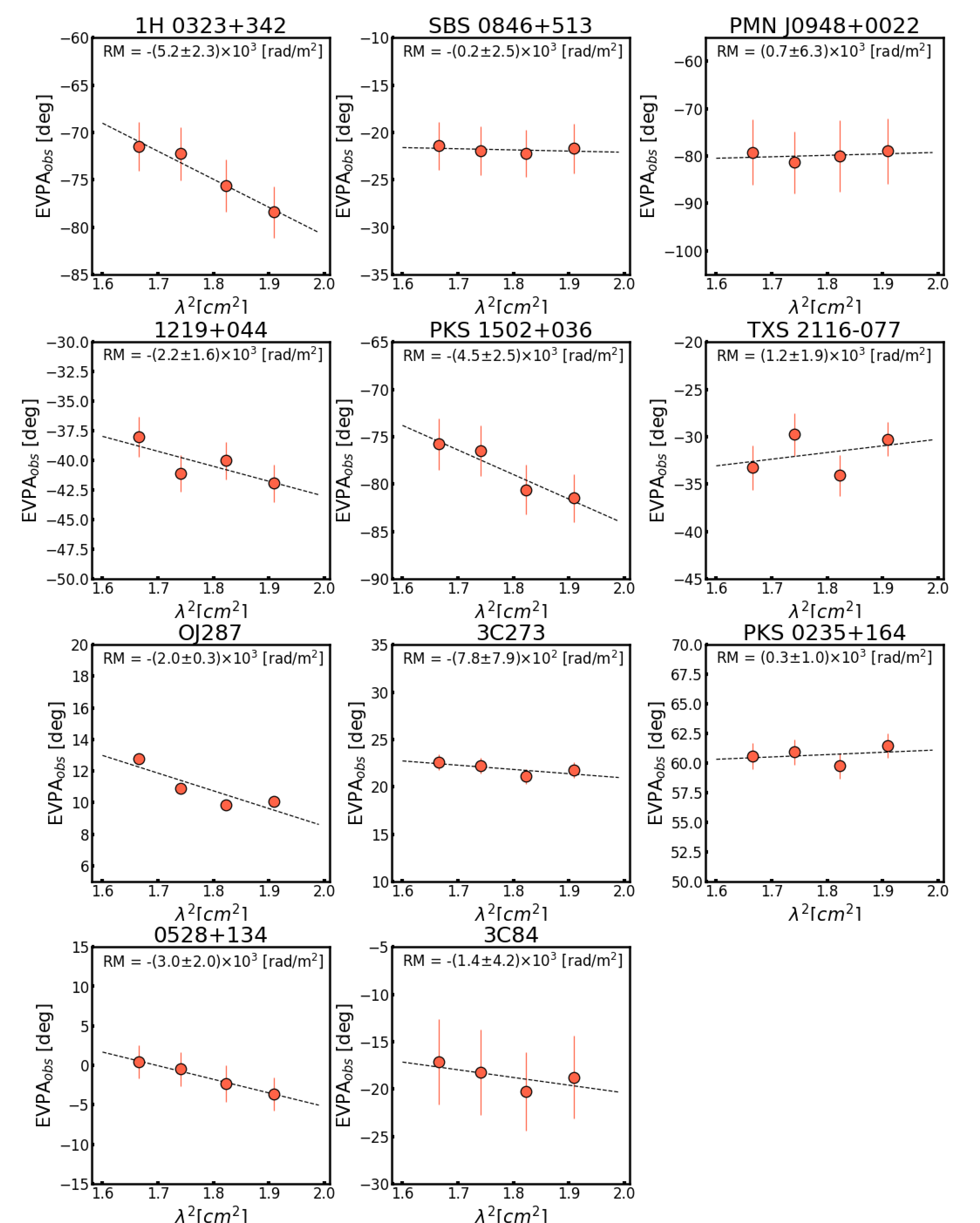}
  \caption{Inband RM results within K band for both NLSy1 targets and calibrators. In each panel, orange filled circles indicate the observed EVPA in each of the four 512\,MHz subbands, and a dashed line indicates a best-fit slope (i.e., RM$_{\rm K}$).}
  \label{fig:inbandrm}
 \end{center}
\end{figure*}

In K band, as we detect sufficient polarization fluxes in all NLSy1 targets, we are able to further produce their polarization images for each of the four 512\,MHz subbands separately. This enables us to examine the RM of detected polarized features over a 2\,GHz bandwidth within K band. In Figure\,\ref{fig:inbandrm}, we show the results of inband RM measurements within K band for all the six targets and five calibrators. Note that no absolute EVPA corrections are performed in these plots (as denoted by EVPA$_{\rm obs}$). Nevertheless, inband RM is determined without correcting absolute EVPA as the four subbands have a common instrumental EVPA offset that does not change the slope in EVPA--$\lambda^2$ plane.
Moreover, the inband RM measurement can avoid $\pm n\pi$ ambiguity of EVPA that is often problematic for the conventional band-to-band RM measurement. 

As expected, inband RM is typically better determined in the calibrators (except for 3C\,84, which is very weakly polarized) than in NLSy1 targets, as the calibrators have sufficiently higher polarization fluxes. The most robust inband RM determination was achieved in OJ\,287, where we detect non-zero RM ($\sim$$2\times10^3\,\rm{rad\,m^{-2}}$) at $>$6$\sigma$. This level of RM is indeed in good agreement with that which was previously reported in the core of OJ\,287 between 22 and 43\,GHz~\citep{Park2018}.  

For NLSy1 targets, the measured inband RM values of SBS\,0846+513, PMN\,J0948+0022 and TXS\,2116-077 are consistent with zero within our 1$\sigma$ uncertainty. Conversely, we detect relatively large RM values ($\sim$ a few times $10^3\,\rm{rad\,m^{-2}}$) in 1H\,0323+342, 1219+044, and PKS\,1502+036 at 1.4--2.3$\sigma$ levels. Remarkably, inband RM is detected in only a few mJy polarization sources, demonstrating the power of wideband observing capability of VERA. The results of inband RM measurement in K band (denoted by RM$_{\rm K}$) are summarized in Table\,\ref{tab:rm}.

\begin{deluxetable*}{ccccccc}
\tablenum{5}
\tablecaption{RM results obtained by VERA wideband polarimetric observations\label{tab:rm}}
\tablewidth{0pt}
\tablehead{
\colhead{Source} &\colhead{Class}&\colhead{$m_{\rm K}$(\%)} & \colhead{$m_{\rm Q}$(\%)}&\colhead{RM$_{\mathrm{K}}$ ($\mathrm{rad\,m^{-2}}$)}&\colhead{RM$_{\mathrm{KQ}}$ ($\mathrm{rad\,m^{-2}}$)}&\colhead{EVPA$\mathrm{_{Q, abs}}$ (deg)} }
\decimalcolnumbers
\startdata
1H 0323+342    &NLSy1 &$2.3\pm0.2$&$1.8\pm0.3$& $-(5.2\pm2.3)\times10^3$&$-(3.3\pm1.2)\times10^3$&$-111.8\pm10.2$\\
SBS 0846+513  &NLSy1&$2.8\pm0.1$&$2.8\pm0.2$& $-(0.2\pm2.5)\times10^3$&$-(3.6\pm0.8)\times10^3$&$-43.7\pm6.5$\\
PMN J0948+0022  &NLSy1&$0.8\pm0.1$&$1.4\pm0.3$& $(0.7\pm6.3)\times10^3$& $(1.9\pm0.9)\times10^3$&$-152.1\pm6.5$\\
PKS 1502+036  &NLSy1&$2.0\pm0.1$&$\leq 7$\textsuperscript{(a)}& $-(4.5\pm2.5)\times10^3$&--&--\\
TXS 2116-077 & NLSy1&$4.2\pm0.1$&$\leq 19$\textsuperscript{(a)}&$(1.2\pm1.9)\times10^3$&--&--\\
1219+044 & NLSy1&$3.1\pm0.1$&$3.4\pm0.4$&$-(2.2\pm1.6)\times10^3$&$-(1.8\pm0.8)\times10^3$&$-78.5\pm6.9$\\
\hline
0528+134   & FSRQ&$2.0\pm0.1$&$3.0\pm0.1$&$-(3.0\pm2.0)\times10^3$&$-(4.3\pm0.6)\times10^3$&$-23.1\pm5.1$\\
3C273      & FSRQ&$4.2\pm0.1$&$1.5\pm0.1$&$-(7.8\pm7.9)\times10^2$&$-(3.3\pm0.5)\times10^3$&$7.4\pm5.3$\\
OJ287   & BLO&$8.5\pm0.1$&$11.0\pm0.1$&$-(2.0\pm0.3)\times10^3$&$-(2.0\pm0.3)\times10^3$&$-29.3\pm4.9$\\
PKS 0235+164  & BLO &$4.0\pm0.1$&$2.9\pm0.1$&$(0.3\pm1.0)\times10^3$ & $-(6.5\pm5.6)\times10^2$&$14.2\pm5.0$\\
3C84   & RG&$0.3\pm0.2$&$0.09\pm0.1$&$-(1.4\pm4.2)\times10^3$& $(2.2\pm0.8)\times10^3$&$-91.6\pm5.7$\\
\hline
\enddata
\tablecomments{(1) Source. (2) NLSy1: Narrow-line Seyfert 1 galaxy; FSRQ: Flat-spectrum radio quasar; BLO: BL Lac object; RG: Radio galaxy (3) Fractional polarization at 22\,GHz. (4) Fractional polarization at 43\,GHz. }\textsuperscript{(a) } An upper limit estimated from the image sensitivity. (5) Inband RM within K band. (6) Band-to-band RM between K and Q bands. (7) Absolute EVPA at Q band.
\end{deluxetable*}

\subsection{Absolute EVPA in Q band}\label{ssec:results-evpa}
Although our VERA data cannot perform absolute EVPA corrections in both K and Q bands by itself, the Boston University Blazar group provides publicly-avaialble fully-calibrated VLBA Q-band polarimetric images of OJ\,287 and PKS\,0235+164 that were observed on 2022 March 12 (a day before our VERA Q-band observations)\footnote{ \url{https://www.bu.edu/blazars/BEAM-ME.html}}. These near-in-time VLBA images of the two calibrators enable us to perform absolute EVPA corrections of our VERA Q-band data by referencing the observed EVPA in VERA images to the fully-calibrated EVPA in VLBA images (using the same polarization features). The EVPA differences of VERA--VLBA were derived to be $53.4\pm0.6^\circ$ and $56.1\pm1.5^\circ$ for OJ\,287 and PKS\,0235+164, respectively. These values derived for two independent calibrators are in good agreement with each other, supporting the validity of our calibration strategy. We adopt weighted mean of $\Delta\chi_{\rm Q} = 53.7\pm0.7^\circ$ for an absolute EVPA correction factor of VERA Q-band data. The absolute Q-band EVPA (EVPA$_{\rm Q, abs}$) derived for each object after the correction is summarized in Table\,\ref{tab:rm}. We will discuss the implications of the Q-band absolute EVPA on our NLSy1 targets in Section~\ref{ssec:discussion-evpa}.

\subsection{RM between K and Q bands}

\begin{figure*}[htbp]
 \begin{center}
  \includegraphics[width=1.0\linewidth]{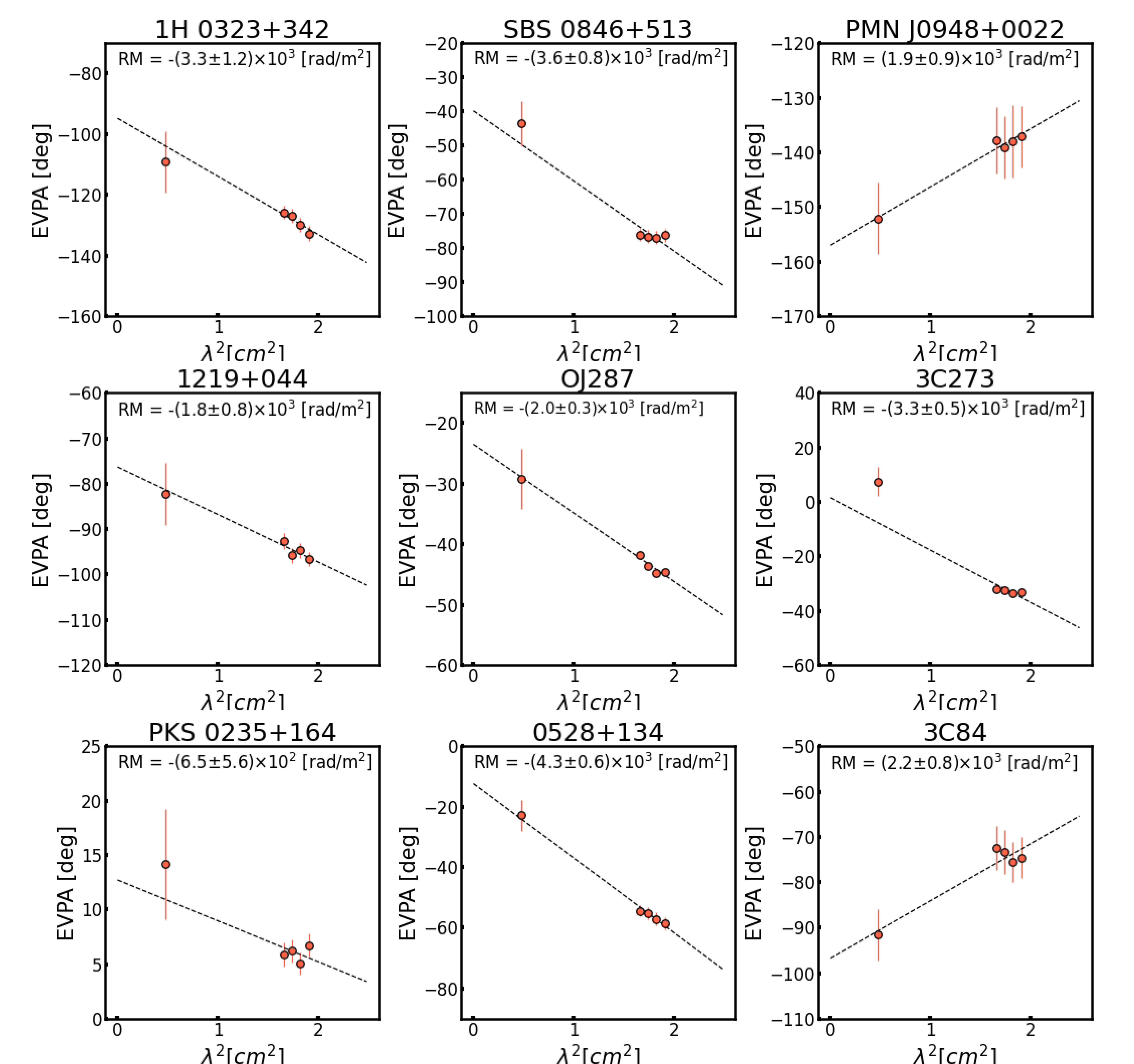}
  \caption{Band-to-band RM results between K and Q bands. Both NLSy1s targets and calibrators are shown. In each panel, EVPAs in K band are obtained in each 512\,MHz subband separately, while in Q band a single EVPA is derived by averaging the whole 2048\,MHz bandwidth. When we measure EVPA at Q band, the beam size of the Q-band image is matched to that of K band images. A dashed line indicates a best-fit slope over the two bands (i.e., RM$_{\rm KQ}$). Absolute EVPA corrections based on Q-band data are performed in the vertical axis (see the main text in detail). }
  \label{fig:kqrm}
 \end{center}
\end{figure*}

Although our VERA inband RM measurement in K band is unique, the weak polarization nature of NLSy1s still leaves a rather large uncertainty on the determination accuracy of their RMs. To better constrain RM with an expanded $\lambda^2$ space, here we additionally make use of Q band data. Our calibration strategy is as follows:

In our present analysis, absolute EVPA corrections are only performed to Q-band data (Section\,\ref{ssec:results-evpa}). In this case, an unknown instrumental EVPA offset still remains in the K-band data. To estimate this offset ($\Delta\chi_{\rm K}$), we needs a reference source for which K-band EVPA relative to that of Q-band is reasonably known. Here we select OJ\,287 as the reference for this purpose because of the following reasons: (1) its inband RM within K band is successfully determined with VERA at a high significance; (2) while some radio sources (such as 3C\,273 and 3C\,84) are suggested to have different RM values in different radio frequency bands~\citep{Hovatta2019, Park2018, Kim2019}, RM of OJ\,287 is known to be relatively constant over a wide range of radio frequencies (22--86\,GHz; \citealt{Park2018}). Therefore, we are able to reasonably connect K-band and Q-band EVPAs (and thus a constant EVPA offset inherent in K band is removed) by extrapolating the slope of RM determined within K band band to Q band. Through this process, we obtained an offset factor $\Delta\chi_{\rm K}=88.0^\circ\pm4.9^\circ$ for OJ\,287 at K band. Subsequently, this correction factor was applied to the whole K-band data to measure the band-to-band RM between K and Q bands for the rest of the sources.

Extended RM plots covering both K and Q bands obtained through the above process is shown in Figure\,\ref{fig:kqrm}. One can see that the RM slopes determined within K band appear to smoothly connect to Q-band EVPA in most of the sources (with the exception of OJ\,287 that is used as the reference), implying that RM does not significantly change over the two frequency bands in these sources. We then perform RM fitting using the whole K and Q band data. For most of the sources, the derived band-to-band RM ($\mathrm{RM_{\mathrm{KQ}}}$) and inband RM ($\mathrm{RM_{\mathrm{K}}}$) are consistent with each other, but now the formal error of $\mathrm{RM_{\mathrm{KQ}}}$ is sufficiently reduced owing to the much wider coverage in the $\lambda^2$ space. Conversely, SBS\,0846+513 and 3C\,273 appear to show rather different $\mathrm{RM_{\mathrm{KQ}}}$ and $\mathrm{RM_{\mathrm{K}}}$, with a tendency of a larger value when Q band is included, implying a possible frequency dependence of RM. We will discuss this in more detail in Section\,\ref{sec:rmdependency}. The results of $\mathrm{RM_{\mathrm{KQ}}}$ are also summarized in Table\,\ref{tab:rm} along with $\mathrm{RM_{\mathrm{K}}}$.

\subsection{Intrinsic RM}

The relationship between the observed $\mathrm{RM}$ and the rest-frame rotation measure $\mathrm{RM_{int}}$ is given as follows:

\begin{equation}
    \mathrm{RM} = \frac{\mathrm{RM_{int}}}{(1+z)^2}
\label{eq:4}
\end{equation}

In Table\,\ref{tab:rmint}, we summarize the observed and redshift-corrected RM for 6 NLSy1s and 5 calibrators. Here we use $\mathrm{RM_{\mathrm{KQ}}}$ for the calculation of $\mathrm{RM_{int}}$, while $\mathrm{RM_{\mathrm{K}}}$ is used for PKS\,1502+036 and TXS\,2116--077, for which $\mathrm{RM_{\mathrm{KQ}}}$ is not available. We find that the magnitudes of $\mathrm{RM_{int}}$ for NLSy1s are as large as $\sim$10$^4$\,rad\,m$^{-2}$ with a mean value of $5.9\times10^3$\,rad\,m$^{-2}$. 

\begin{deluxetable}{cccc}
\tablenum{6}
\tablecaption{Summary of redshift $z$ and $\mathrm{RM_{int}}$\label{tab:intrm}}
\tablewidth{0pt}
\label{tab:rmint}
\tablehead{
\colhead{Source} &\colhead{z}&\colhead{$\mathrm{RM}$}&\colhead{$\mathrm{RM_{int}}$}\\
\colhead{}&\colhead{}&\colhead{$(\times10^3\,\mathrm{rad/m^2})$}&\colhead{$(\times10^3\,\mathrm{rad/m^2})$} }
\decimalcolnumbers
\startdata
1H\,0323+342 & 0.061 &$-(3.3\pm1.2)$& $-(3.7\pm1.4)$ \\
SBS\,0846+513    & 0.584 &$-(3.6\pm0.8)$& $-(9.0\pm2.0)$ \\
PMN\,J0948+0022    & 0.585 &$(1.9\pm0.9)$& $(4.8\pm2.3)$ \\
1219+044   & 0.966 &$-(1.8\pm0.8)$& $-(7.0\pm3.1)$ \\
PKS 1502+036&0.408&${-(4.5\pm2.5)}$\textsuperscript{(a)}&$-(8.9\pm5.0)$\\
TXS 2116-077&0.26&${(1.2\pm1.9)}$\textsuperscript{(a)}&$(1.9\pm3.0)$\\
\hline
0528+134 &2.06 &$-(4.3\pm0.6)$&$-(40\pm6)$\\
3C273 & 0.158&$-(3.3\pm0.5)$&$-(4.4\pm0.7)$\\
OJ287 &0.31&$-(2.0\pm0.3)$&$-(3.4\pm0.5)$ \\
PKS 0235+164 &0.94&$-(0.65\pm0.56)$& $-(2.4\pm2.1)$\\
3C84 &0.0176&$(2.2\pm0.8)$&$(2.3\pm0.8)$\\
\hline
\enddata
\tablecomments{(1) Source name. (2) Redshift. (3) Band-to-band RM results at 22/43\,GHz (RM$_{\rm KQ}$). \noindent{\footnotesize{\textsuperscript{(a)} Inband RM results (RM$_{\rm K}$).}} (4) Intrinsic ($z$-corrected) RM.}
\end{deluxetable}

\subsection{Notes on individual target sources}\label{ssec:individual}
\subsubsection{1H 0323+342}
1H\,0323+342 is known as the nearest ($z=0.061$) $\gamma$-ray detected NLSy1 \citep{Abdo2009b}, and one of the very few NLSy1s where the host galaxy is imaged \citep{Zhou2007, Leon2014}. The central BH mass of this galaxy has been estimated to be $\sim$$2\times 10^7\,M_{\odot}$ by various methods \citep{Landt2017, Zhou2007,Wang2016}, although a possibility of a larger BH mass is also debated \citep{Leon2014, Hada2018, D'Ammando2020}. 1H\,0323+342 is a unique target that enables us to probe the vicinity of the
central engine at the highest linear resolution among $\gamma$-ray detected NLSy1s (i.e., 1\,mas = 1.2\,pc). Previous high-dynamic-range VLBA observations revealed a well-collimated one-sided jet structure toward the southeast at (sub)parsec scales~\citep{Wajima2014,Hada2018} with highly superluminal features up to $\sim$9$c$~\citep{Lister2016,Fuhrmann2016,Doi2018}.
In our VERA K/Q-band polarimetric observations, we detect a significant polarization feature in the central region but its peak location appears to be slightly ($\sim$0.3\,mas) offset from the peak position of the total intensity. The amount of polarized flux on the jet feature located at $\sim$ 7 mas is quite small so that we were not able to detect the polarized flux in each sub-band at K band and obtain inband RM in this region.

\subsubsection{SBS\,0846+513}
This source is one of the well-studied $\gamma$-ray emitting NLSy1s. Large-amplitude $\gamma$-ray flares have been observed from this source in 2011 \citep{D'Ammando2012, D'Ammando2013}. The central BH mass of this source has also been under discussion, with values in a range of $(2.5-6.4)\times10^8\,M_{\odot}$ \citep{D'Ammando2020}. 
Previous high-resolution VLBA observations at $\leq$15 GHz clearly resolved a core-jet structure in the east-west on parsec scales. SBS\,0846+513 is the first NLSy1 where a superluminal motion was measured, suggesting the existence of a highly beamed relativistic jet \citep{D'Ammando2012}.  An optical flare with a high optical fractional polarization ($\sim$10\%) reported in 2013 also supports the blazar-like nature of this source~\citep{Maune2014}. Our VERA observations in K/Q bands detect intense polarization fluxes in the core region. Interestingly, while EVPAs are relatively constant across K band, we observe a significant difference in EVPA by $\sim$30$^{\circ}$ between K and Q bands, suggesting a possible frequency dependence in RM.

\subsubsection{PMN\,J0948+0022}
This source is the most radio-loud NLSy1; moreover, it is the first NLSy1 where $\gamma$-ray emission was detected with \textit{Fermi}-LAT~\citep{Abdo2009a}. During 2012 December -- 2013 
January the source exhibited the most powerful flaring activity from optical to $\gamma$-rays \citep{D'Ammando2015}. Previous high-resolution VLBI observations revealed a very compact structure with a high brightness temperature on parsec scales~\citep{Doi2006}, with a slight extension toward the north-east direction~\citep{Giroletti2011}. The X-ray spectral slope is similar to that found in radio-quiet NLSy1s, suggesting that a standard accretion disk is present as expected from its high accretion rate~\citep{D'Ammando2014}. 
In our VERA observations, the parsec-scale structure is essentially unresolved with the highest core flux density among our NLSy1 targets ($\sim$500\,mJy at K band). In contrast, the core fractional polarization is found to be the lowest ($m_{\rm K}$$\sim$0.8\%) in our NLSy1 sample. \citet{Giroletti2011} reported a similar level of core fractional polarization ($\sim$0.9\% at 22\,GHz) with European VLBI Network observations in 2010 June, while \citet{Foschini2011a} reported an enhanced polarization flux with VLBA at 15\,GHz during an outburst phase in 2010 September.

\subsubsection{PKS\,1502+036}
PKS\,1502+036 is a powerful radio-loud NLSy1 with its $\gamma$-ray emission detected in the early phase of  $Fermi$-LAT observations~\citep{Abdo2009b}.  The source is highly variable and the broadband SED fits well with a model of synchrotron radiation plus inverse Compton scattering, suggesting that its high-energy emission is of jet origin rather than due to starbursts \citep{Richads2011,D'Ammando2016}. \citet{Yuan2008} estimated a central BH mass of this galaxy to be $4\times10^6\,\mathrm{M_{\odot}}$ by the virial method. However, near-infrared bulge luminosity of the host galaxy survey and modeling of the optical-UV spectra with a standard thin accretion disk obtained a BH mass of $\sim (3-7) \times10^8\,M_{\odot}$~\citep{D'Ammando2018, Calderone2013}. Previous VLBA observations of PKS\,1502+036 revealed a core-jet structure on parsec scales, but no apparent superluminal motion was detected during 2008--2012~\citep{D'Ammando2013}. The MOJAVE program reported low levels of core fractional polarization at 15\,GHz~\citep{Hodge2018}. The present VERA observations for the first time detect core polarization and RM of this source up to K band. However, we had instrumental issues during the scans of this source in our Q-band session, resulting in non-detection of polarization signals due to the considerbly poorer sensitivity (Table\,\ref{tab:imageparameters}).

\subsubsection{TXS\,2116-077}
The nature of the source is still debated. This source has originally been classified as a $\gamma$-ray emitting NLSy1 in the 4FGL catalog. However, new high-quality spectroscopic data of this source disfavor its classification as a NLSy1 \citep{Jarvela2020}, suggesting to be an intermediate-type Seyfert galaxy. This source is hosted in a merging system and its companion galaxies both have pseudo-bulges and are late-type galaxies~\citep{Jarvela2020}. Previous VLA observations at 8.4 GHz indicated a very compact morphology on subarcsecond scales~\citep{Paliya2018}. More recently, high-resolution VLBA observations at 2--15\,GHz have revealed a core-jet morphology (towards the south) on mas scales together with significant core flux variability~\citep{Shao2023}, the characteristics of which are similar to blazars. Our VERA observations for the first time detect the parsec-scale structure of this source up to 22 and 43\,GHz. The total-intensity morphology is essentially unresolved at both frequencies, and the detected polarization emission at 22\,GHz is coincident with the core. Similarly to PKS\,1502+036, we had instrumental issues during the scans of this source in the Q-band session, resulting in non-detection of polarization signals in this band.

\subsubsection{1219+044}
Although 1219+044 is classified as a FSRQ in the 4FGL catalog, this source has been suggested to be a NLSy1 candidate by \citet{Kynoch2019}. Measurements of optical emission lines for this source give values below or above $2000\,\mathrm{km\,s^{-1}}$, depending on the line shape and the analysis technique used, making its classification as a NLSy1 debated. For this reason, we consider this source as a candidate NLSy1. At a redshift of 0.966 (1\,mas = 8.15\,pc, \citealt{Paris2017}), 1219+044 is the most distant target in our sample.
Previous VLBI observations at 15--43\,GHz revealed a highly core-dominated mas-scale structure with a small amount of jet-like extended emission towards the south \citep{Lister2019, Cui2021}.
In our VERA images, the structure is essentially unresolved in both K and Q bands. Significant polarization fluxes are detected in the position coincident with the unresolved core.

\section{Discussion}
\subsection{RM versus optical classification}


Probing the nuclear environment of AGNs is crucial for understanding their diverse accretion/ejection activities and possible link between different AGN classes. Parsec-scale RM towards the nucleus provides a clue to this since it is directly related to the key physical quantities of the Faraday screen surrounding the nucleus such that $\mathrm{RM} = 8.1 \times 10^{5} \int\left(\frac{n_{\mathrm{e}}}{\mathrm{cm}^{-3}}\right)\left(\frac{B_{\|}}{G}\right)\left(\frac{\mathrm{d} l}{\mathrm{pc}}\right)\,\mathrm{rad\,m^{-2}}$, where $n_e$, $B_{\|}$ and $l$ are the electron density, strength of magnetic field parallel to the line-of-sight, and path length of the Faraday screen, respectively. There are various potential sites of external Faraday rotation in radio-loud AGN: immediate surroundings of the radio jet such as winds/outflows/jet sheath \citep[e.g.,][]{Wardle2018}, broad-line regions at (sub)parsec scales, narrow-line regions far from the nucleus \citep[e.g,][]{Taylor1998} as well as Faraday rotation internal to the jet~\citep[e.g.,][]{Homan2012}, whereas it is practically challenging to identify the dominant source of Faraday rotation exactly. Nevertheless, RM generally serves as a useful probe to assess the physical conditions around a SMBH.

So far, parsec-scale core RM of radio-loud AGNs has largely been examined for highly beamed sources, i.e., FSRQs and BLOs. \citet{Zavala2003} first noticed that the core RMs of FSRQs tend to be larger than those of BLOs based on their VLBA 8--15\,GHz survey for 40 sources. \citet{Hovatta2012} later obtained a consistent trend for a larger number of samples in the same frequency range. More recently, such a trend has also been indicated at higher frequencies~\citep[22--86\,GHz;][]{Park2018}. The larger core RM values in FSRQs than in BLOs are in accordance with the standard picture that FSRQs have more material around them, resulting in the higher accretion luminosity and more powerful outflows.

\begin{figure*}[htbp]
\centering \includegraphics[scale=0.3]{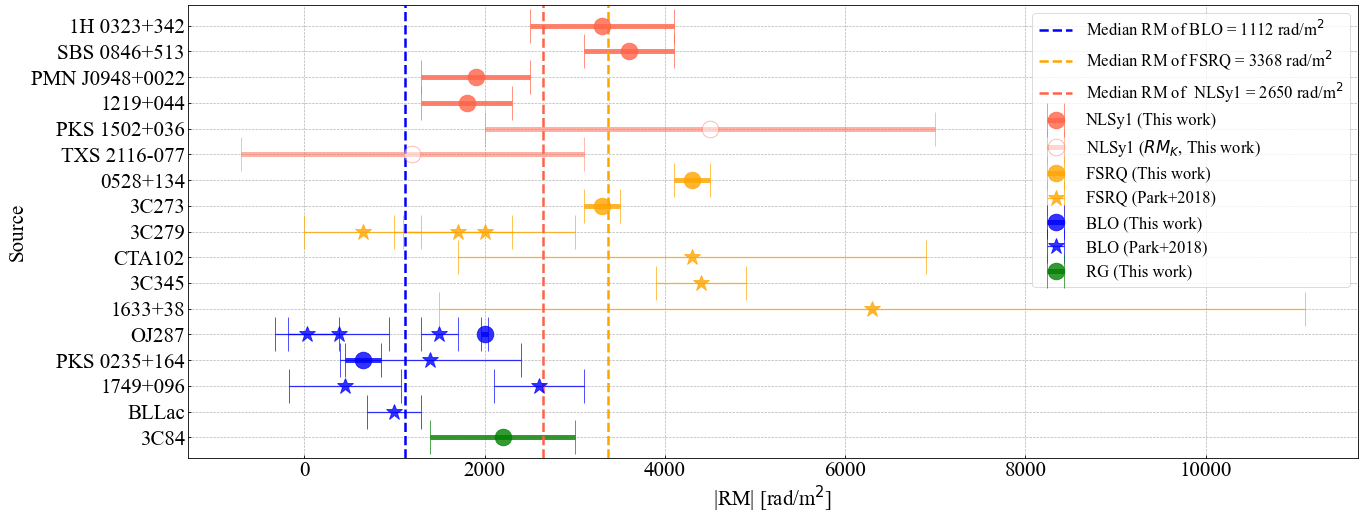}
\caption{Observer-frame core RM distribution on various jetted AGNs. RM values measured between K and Q bands are presented except for PKS\,1502+036 and TXS\,2116-077, for which inband RM values within K band are used instead. Different color codes are used for different AGN classes. To increase the sample of FSRQs/BLOs, the literature results of RM$_{\rm KQ}$ obtained with KVN~\citep{Park2018} are also added. Circle and star symbols are from this work and ~\citet{Park2018}, respectively. Red/orange/blue-colored vertical dashed lines represent median RM values for NLSy1s/FSRQs/BLOs, respectively.}
\label{fig:RMdistribution}
\end{figure*}

Along with this line, now it would be intriguing to compare the observed core RM of RL-NLSy1s with those of FSRQs/BLOs. As the six NLSy1s examined in this paper are all high beamed sources similar to FSRQs/BLOs, any differences/similarities in RM properties would reflect the intrinsic difference/similarity among them rather than the orientation effect. 
In Figure\,\ref{fig:RMdistribution}, we compare the observed magnitude of RM$_{\rm KQ}$ (or RM$_{\rm K}$ where RM$_{\rm KQ}$ is not available) for all our targets/calibrators. To increase the sample of FSRQs/BLOs, here we add the results of RM$_{\rm KQ}$ obtained with KVN by \citet{Park2018}. We are able to confirm the previously-suggested trend that the core RM values of FSRQs (a median of $3.4\times10^3\,\mathrm{rad\,m^{-2}}$) are larger than those of BLOs (a median of $1.1\times10^3\,\mathrm{rad\,m^{-2}}$). For RL-NLSy1s, interestingly, their core RM values appear to be distributed in between, with a median of $2.7\times10^3\,\mathrm{rad\,m^{-2}}$. A Kolmogorov-Smirnov (KS) test finds a probability of 3\% and 4\% that BLO and FSRQ, and BLO and RL-NLSy1 are from the same parent population, respectively. Conversely, FSRQ and RL-NLSy1 have a 51\% probability of coming from the same population. If we consider the result significant when the probability is $\leq$5\%, we can say that the RM properties of RL-NLSy1s are intrinsically different from those of BLOs, while rather similar to those of FSRQs. If we examine this using the redshift-corrected RM (RM$_{\rm int}$), the results of the KS tests are essentially unchanged, but the differences of characteristic RM$_{\rm int}$ values among the three classes becomes more apparent, with median RM$_{\rm int}$ values of $2.3\times10^3$, $6.1\times10^3$, and $1.7\times10^4$\,$\mathrm{rad\,m^{-2}}$ for BLO, RL-NLSy1, and FSRQ, respectively (see Figure\,\ref{fig:intRMdist}).

\subsection{Anticorrelation between core RM and $m$}
\begin{figure}[h]
\centering \includegraphics[scale=0.58]{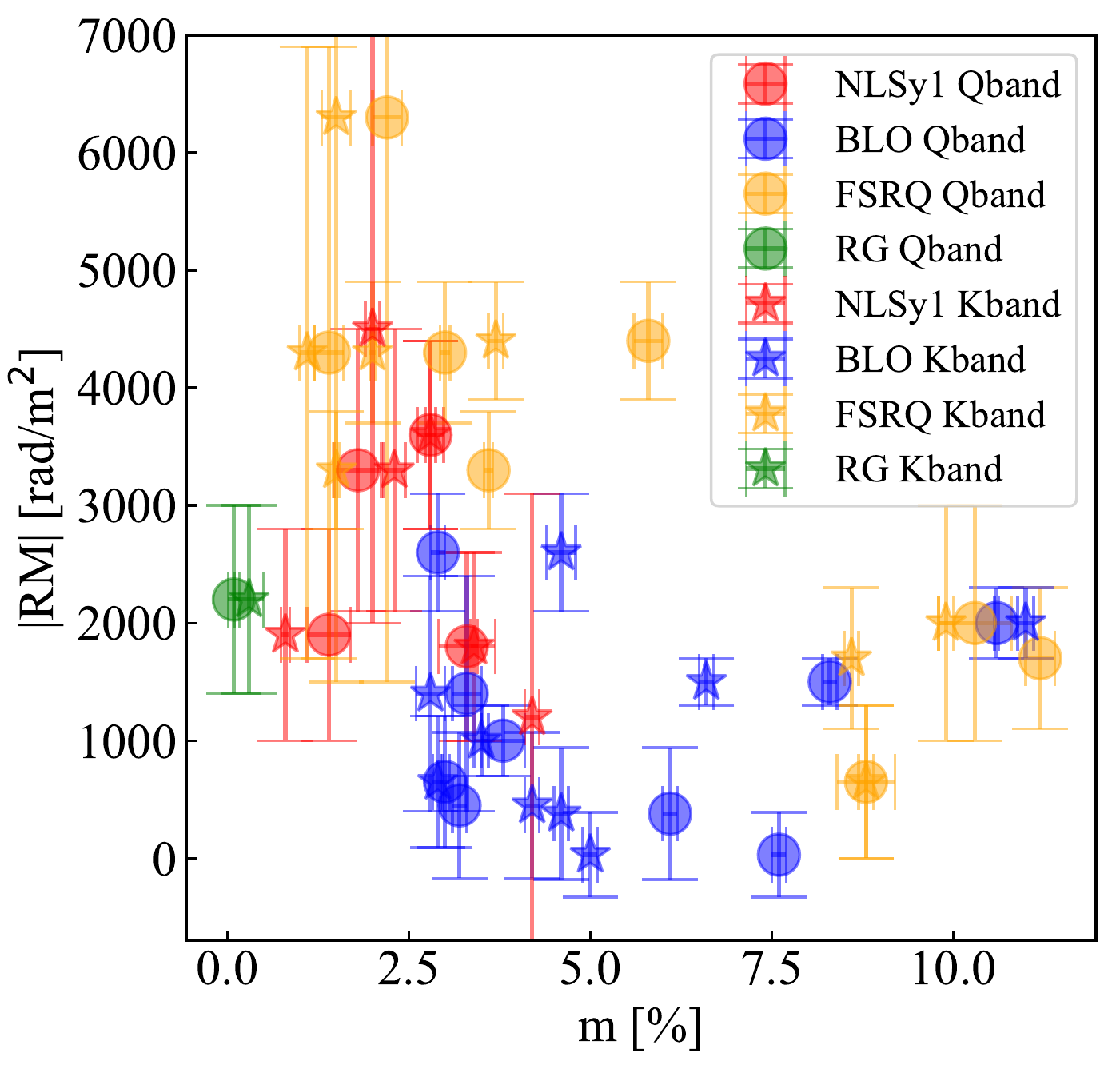}
\caption{Observer-frame core RM$_{KQ}$ magnitude versus core fractional polarization degree at K (star symbols) and Q (circle symbols) bands. Plotted data are taken from Table \ref{tab:rm} and \citet{Park2018}. Similarly to Figure\,\ref{fig:RMdistribution}, PKS\,1502+036 and TXS\,2116--077 data are based on the Inband RM results at K band.}
\label{fig:coreRM}
\end{figure}

\citet{Zavala2003} and \citet{O'Sullivan2009} found that the parsec-scale core RM of blazars is anticorrelated with the core degree of polarization at $\leq$15\,GHz, in the sense that higher core RM sources tend to show lower core fractional polarization $m$. \citet{Zavala2003} interpreted the $\mathrm{RM}$--$m$ anticorrelation by Faraday depolarization within the observed band due to large RM and/or due to large spatial RM gradient of Faraday screen within the observing beam size. Here we revisit this issue by adding our RM results obtained for RL-NLSy1s. Figure\,\ref{fig:coreRM} shows the plot of the magnitude of the observed core RM for our sample as a function of $m$ measured at K and Q bands, together with the previous core RM results on FSRQs/BLOs obtained in the same frequency bands~\citep{Park2018}. We can see that the trend of $\mathrm{RM_{int}}$--$m$ anticorrelation reported at the low frequencies is maintained also at the high frequencies, where FSRQs dominate the high RM and low $m$ space while BLOs preferably dominate the low RM and high $m$ space. 
Note that an FSRQ 3C279 (taken from Park et al. 2018) is placed in the space of low RM and high m, which is apparently different from the rest of FSRQs. However, the measurement by Park et al. 2018 was made by KVN with large beam sizes and the core fractional polarization of 3C279 was likely contaminated by the bright inner jet with higher fractional polarization\citep{Park2018}.

RL-NLSy1s tend to fall into the space somewhere between FSRQs and BLOs, largely following the trend of RM--$m$ anticorrelation. If we examine this using $\mathrm{RM_{int}}$--$m$ instead of RM--$m$, the anticorrelation appears to be even more prominent (see Figure\,\ref{fig:coreintRM}). Hence, the updated RM-$m$ plot with the addition of RL-NLSy1s remains consistent with the scenario that the larger Faraday rotation leads to the larger depolarization. Although our results cannot identify the exact cause of depolarization by themselves, the presented RM--$m$ result additionally supports that the nuclear regions of RL-NLSy1 are more Faraday-dense than those of BLOs.

\subsection{Frequency dependence of RM}\label{sec:rmdependency}
Although traditional measurement of RM determines a single RM value over multiple radio bands \citep[e.g.,][]{Hovatta2012,Zavala2003}, some recent polarimetric studies of AGN jets indicate that the magnitude of RM varies with frequency in the sense that RM values measured in higher frequency bands are systematically larger than those measured in lower frequency bands \citep{Park2018, Hovatta2019}. As far as FSRQs and BLOs are concerned, our RM values obtained at K/Q bands are in fact significantly larger than the typical RM values obtained in previous studies made at $\leq$15\,GHz \citep{Asada2002, Zavala2001, Zavara2005,Hovatta2012}. 

\begin{figure}[h]
\centering \includegraphics[scale=0.35]{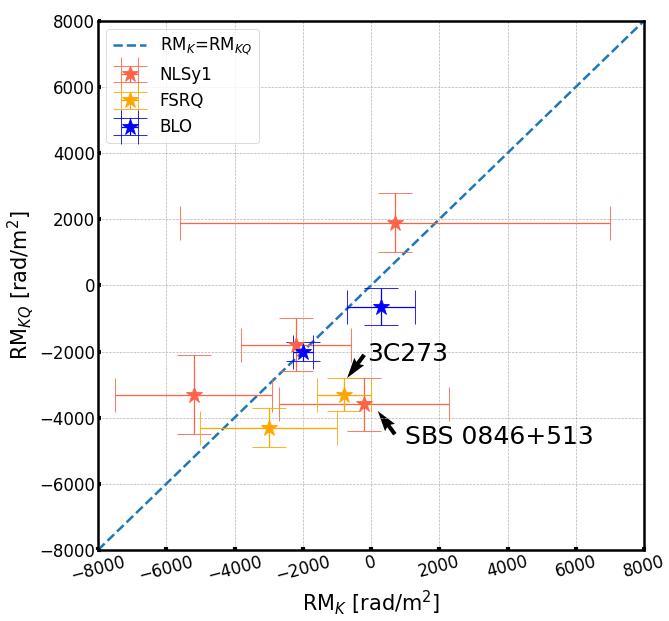}
\caption{$\mathrm{RM_{K}}$ versus ${\mathrm{RM_{KQ}}}$ for the targets and calibrators observed in this study. The blue dashed line corresponds to $\mathrm{RM_{K}}$=${\mathrm{RM_{KQ}}}$.}
\label{fig:rmk-vs-rmkq}
\end{figure}

Here, we additionally attempt to search for the possible frequency dependence of RM by comparing RM determined within K band (${\rm RM_{K}}$) and RM determined between K and Q bands (${\rm RM_{KQ}}$). In Figure\,\ref{fig:rmk-vs-rmkq}, we show a plot of ${\rm RM_{K}}$-versus-${\rm RM_{KQ}}$ for eight sources where both ${\rm RM_{K}}$ and ${\rm RM_{KQ}}$ were measured (four NLSy1s and four calibrators excluding OJ\,287 that was assumed to be ${\rm RM_{K}} = {\rm RM_{KQ}}$ in our EVPA calibration). Although the data points are rather scattered mainly due to the larger uncertainty on ${\rm RM_{K}}$, six out of the eight sources are consistent with ${\rm RM_{K}} = {\rm RM_{KQ}}$ within 1$\sigma$ uncertainty, indicating that frequency dependence is not so obvious in these sources between K and Q bands. Conversely, SBS\,0846+513 (NLSy1) and 3C\,273 (FSRQ) have possible larger magnitudes of ${\rm RM_{KQ}}$ than those of ${\rm RM_{K}}$, implying an increasing RM with frequency. This trend is consistent with that suggested in \citet{Park2018} and \citet{Hovatta2019}. Nevertheless, a better statistics (especially for inband RM) would be needed to more robustly examine this.

\subsection{EVPA vs. Jet position angle}\label{ssec:discussion-evpa}
Here, we briefly discuss the relationship between the jet position angle (PA) and intrinsic EVPA (EVPA$_{\rm int}$) of NLSy1s. We estimate EVPA$_{\rm int}$ of each target by extrapolating the observed RM$_{\rm KQ}$ to $\mathrm{\lambda}^2 = 0$ in Figure\,\ref{fig:kqrm}. As EVPA$_{\rm int}$ can be closely associated with the direction of magnetic field in the emitting plasma, a comparison of EVPA$_{\rm int}$ with the spatially-resolved jet morphology may provide information regarding the internal structure of relativistic jets as well as the associated magnetic field. In Table\,\ref{tab:jet-vs-evpa}, we summarize PA the parsec-scale jet, Q-band absolute EVPA and intrinsic EVPA for four NLSy1 sources where Q-band polarization signals were detected. We find that EVPA$_{\rm int}$ and jet PA are significantly offset with each other in most of NLSy1s, although PMN\,J0948+0022 is rather uncertain due to its extremely compact structure and relatively large error of EVPA. A similar trend was also reported for a sample of NLSy1s in a previous polarimetric study using MOJAVE at 15\,GHz~\citep{Hodge2018}. Interestingly, such a jet-EVPA misalingment appears to be in contrast to that of BLOs, for which parsec-scale EVPAs tend to be aligned with the jet PA; rather similar to those of FSRQ where jet-EVPA is skewed toward misalignments ~\citep{Hodge2018}. This implies that simple transverse shocks (that compress the magnetic field perpendicular to the jet) are less dominant in the jet of NLSy1s. Large jet-EVPA offsets may be caused by various physical reasons such as an additional Faraday screen to the jet base, complex topology of the underlying magnetic field (e.g., helical field), velocity gradients in the jet (e.g., multi-layered structure) and temporal structural variability. In fact, a previous polarization monitoring study for a sample of RL-NLSy1s at optical wavelengths revealed the variable nature of NLSy1's  EVPA~\citep{Angelakis2018}. Future coordinated multi-epoch and multi-wavelength polarimetric observations will help better constrain the intrinsic magnetic-field structure of NLSy1 jets. 

\begin{deluxetable}{cccc}
\tablenum{7}
\tablecaption{Jet PA and intrinsic EVPA for NLSy1s \label{tab:jet-vs-evpa}}
\tablewidth{0pt}
\tablehead{
\colhead{Source} &\colhead{Jet PA}&\colhead{EVPA$_{\mathrm{Q, abs}}$}&\colhead{EVPA$_{\mathrm{int}}$}}
\decimalcolnumbers
\startdata
1H\,0323+342 & 130 \textsuperscript{(a)} & 68$\pm$10 & 85$\pm$12\\
SBS\,0846+513 &115 \textsuperscript{(a)} & 136$\pm$6.5 & 140$\pm$8.0 \\
PMN\,J0948+0022    & 35.4 \textsuperscript{(b)}& 28$\pm$6.5 &23$\pm$8.8\\
1219+044   & $\sim180$ \textsuperscript{(c)} & 102$\pm$6.9 & 104$\pm$8.2\\
\hline
\enddata
\tablecomments{(1) Source name. (2) Jet position angle at 43\,GHz except for PMN\,J0948+0022 (measured from a 22\,GHz image). (\textsuperscript{(a)}Position angles of the major axes of the Gaussian model taken from our 43\,GHz observation}, \textsuperscript{(b)}\citealp{Giroletti2011}, \textsuperscript{(c)}\citealp{Cui2021}). (3) Absolute EVPA at 43\,GHz. (4) Intrinsic EVPA estimated by using RM$_{\mathrm{KQ}}$ and EVPA$_{\mathrm{Q, abs}}$.
\end{deluxetable}
\subsection{RL-NLSy1: a scale-down (younger) system of FSRQ?}\label{ssec:nls-vs-fsrq}

As discussed in the previous subsections, we found that the parsec-scale polarimetric properties of RL-NLSy1s are rather different from those of BLOs, while sharing a number of similarities to FSRQs summarized as follows: 

\begin{itemize}
    \item High core RM values
    \item Low core fractional polarization
    \item Possible increase of core RM at higher frequencies
    \item Misalignment between EVPA and jet PA
\end{itemize}

This is in accordance with the standard AGN model where the nuclear environment of Seyfert galaxies and FSRQs are more gas-rich than those of BLOs. Conversely, we do see a small difference between RL-NLSy1s and FSRQs as the core RM of NLSy1 tends to be smaller than that of FSRQ. What does it imply?  

According to recent studies of AGN evolutionary sequences, NLSy1s are suggested to be in an early stage of AGN evolution~\citep[e.g.,][]{Mathur2000} with their central BH masses being smaller than those of more evolved quasars. For jetted AGN, \citet{Foschini2017} noted the similarity between radio-loud (or $\gamma$-ray detected) NLSy1s and FSRQs with only one key scaling parameter being $M_{\rm BH}$, and proposed an interesting idea that RL-NLSy1s would eventually evolve into FSRQs. Hence, here we consider a toy model that the whole system of RL-NLSy1s (including the Faraday screen) is simply a $M_{\rm BH}$-scale-down version of FSRQs. 
In this case, our pc-scale observations of NLSy1s likely trace the regions more distant from SMBH than in FSRQs in terms of gravitational radius (for sources at the same redshift) as the angular resolution of our observations are limited. If the physical quantities of the Faraday screen are scaled by $B\propto (r/r_g)^{-1}$, $n_{e}\propto (r/r_g)^{-1}$ and $l\propto r/r_g$, where $r_g$ is the gravitational radius $GM_{\rm BH}/c^2$ of SMBH, the radial dependence of RM approximately results in $\mathrm{RM}\propto (r/r_g)^{-1}$. Therefore, RM measured at a given linear distance $r$ results in $\mathrm{RM}\propto M_{\rm BH}$. Such a positive RM--$M_{\rm BH}$ correlation is qualitatively consistent with what we observed in NLSy1s and FSRQs. Of course, this is a very rough and simplified estimate and the exact radial profiles of $B$, $n_{e}$ and $l$ can be more complicated. Nevertheless, unless a very large path length $l$ is required for NLSy1s, a smaller RM in NLSy1s than in FSRQ may be a natural consequence that $M_{\rm BH}$ of NLSy1s are in fact to some extent smaller than that of FSRQs, providing an additional support that NLSy1s and FSRQs are closely linked with each other in terms of AGN evolutionary sequences.

\section{Summary}
We reported the detailed polarimetric observations for the parsec-scale nuclear regions of four bona-fide and two candidate active RL-NLSy1 in K and Q bands. Owing to the wideband recording capability of VERA, we were able to study the weak polarimetric properties of these nuclei in great detail. We summarize our main results and implications as follows: 

\begin{enumerate}
    \item We obtained the parsec-scale total-intensity images at 22 and 43\,GHz for all the six NLSy1 targets. Most of them are compact on parsec scales with a high brightness and flat-to-steep spectral radio core, in agreement with the scenario that $\gamma$-ray detected RL-NLSy1s are highly beamed objects similar to blazars. 
    
    \item We detected significant polarimetric signals in all the six targets (except for 43\,GHz polarization in PKS\,1502+036 and TXS\,2116-077) that are largely coincident with their total-intensity peak. The measured core fractional polarizations $m$ of NLSy1s were all rather modest (no larger than 5\%). 
    
    \item For the first time, we unambiguously detected Faraday rotation towards the parsec-scale radio core of NLSy1s. The rotation measure was determined either within a 2\,GHz passband in K band ($\mathrm{RM_{\mathrm{K}}}$) or between K and Q bands ($\mathrm{RM_{\mathrm{KQ}}}$).  
    
    \item The observed median core RM of NLSy1s was $2.7\times10^3\,\mathrm{rad\,m^{-2}}$ (or $6.1\times10^3\,\mathrm{rad\,m^{-2}}$ for redshift-corrected), which is significantly larger than that of BLOs and similar to or slightly lower than that of FSRQ. We also found that core RM and fractional polarization was anticorrelated in the sense that FSRQs/BLOs preferably dominate the (high RM, low $m$)/(low RM, high $m$) space, while NLSy1s tend to fall into the space between FSRQ and BLO. This is consistent with the previously-suggested scenario that the large Faraday rotation leads to a larger depolarization.  

    \item Comparing $\mathrm{RM_{\mathrm{K}}}$ (inband RM)  with $\mathrm{RM_{\mathrm{KQ}}}$ (band-to-band RM), seven out of nine sources showed $\mathrm{RM_{\mathrm{K}}} = \mathrm{RM_{\mathrm{KQ}}}$ within our uncertainty. Conversely, SBS\,0846+513 (NLSy1) and 3C273 (FSRQ) showed a hint of $\mathrm{RM_{\mathrm{KQ}}} > \mathrm{RM_{\mathrm{K}}}$, implying an increase of RM at higher frequencies.
    
    \item We found a significant misalignment between the jet position angle and intrinsic (RM-corrected) EVPA in NLSy1s, similar to what was previously reported in FSRQs. 

    \item Overall, the revealed parsec-scale polarimetric properties of NLSy1s are significantly different from those of BLOs, despite being similar to those of FSRQs. This is in accordance with the standard AGN picture that the nuclear environment of NLSy1s and FSRQs are more gas-rich than those of BLOs. Furthermore, smaller core RM values of NLSy1s than of FSRQ could be explained by smaller central BH masses in NLSy1s, supporting the scenario that NLSy1s are in an early stage of AGN evolution. 

\end{enumerate}

We also summarize our future prospects as follows:

\begin{itemize}
    \item {\em RM variability}: While we successfully obtained the core RM on six RL-NLSy1s, it is still uncertain whether RM in these objects are time variable or not. Variability time scales of RM provide a critical insight about the spatial extent of the Faraday screen~\citep[e.g.,][]{Wardle2018}, which in turn will allow us to better constrain the nuclear environment and the exact origin of RM in these NLSy1s.
    \item {\em Other RL-NLSy1s}: While the present study focused only on the most prominent $\gamma$-ray emitting RL-NLSy1s as a pilot project, they are rare examples among the whole NLSy1 population. To better understand the general properties of polarization and RM of (active) NLSy1s, it is important to extend our sample to other $\gamma$-ray emitting NLSy1s and also non-$\gamma$-ray emitting radio-loud NLSy1s. 
    \item {\em Higher frequency observations}: While the present VERA study was limited to 22 and 43\,GHz, higher frequency observations will be able to probe the closer vicinity of the central SMBH thanks to the less synchrotron opacity to the radio core. A combined analysis of VERA with the data obtained from high frequency radio facilities, such as KVN (up to 129\,GHz) and the the Atacama Large Millimeter/submillimeter Array (ALMA), will enable us to reveal the radial evolution of the nuclear environment of NLSy1s.
    
\end{itemize}

\begin{figure*}[htbp]
 \begin{minipage}{0.1\hsize}
  \begin{center}
   \includegraphics[width=45mm]{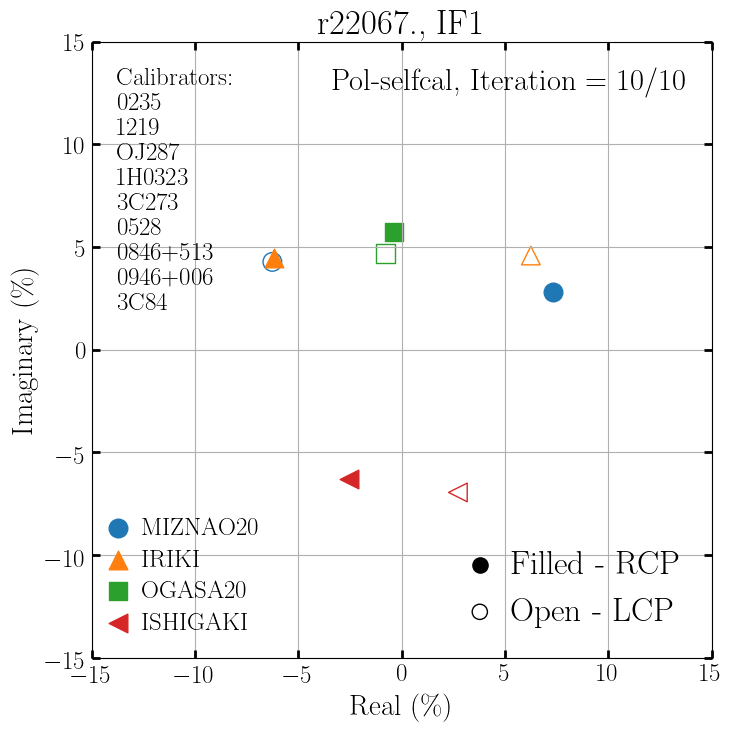}
  \end{center}
  \label{fig:one}
 \end{minipage}
 \hspace{+0.3\columnwidth}
 \begin{minipage}{0.1\hsize}
 \begin{center}
  \includegraphics[width=45mm]{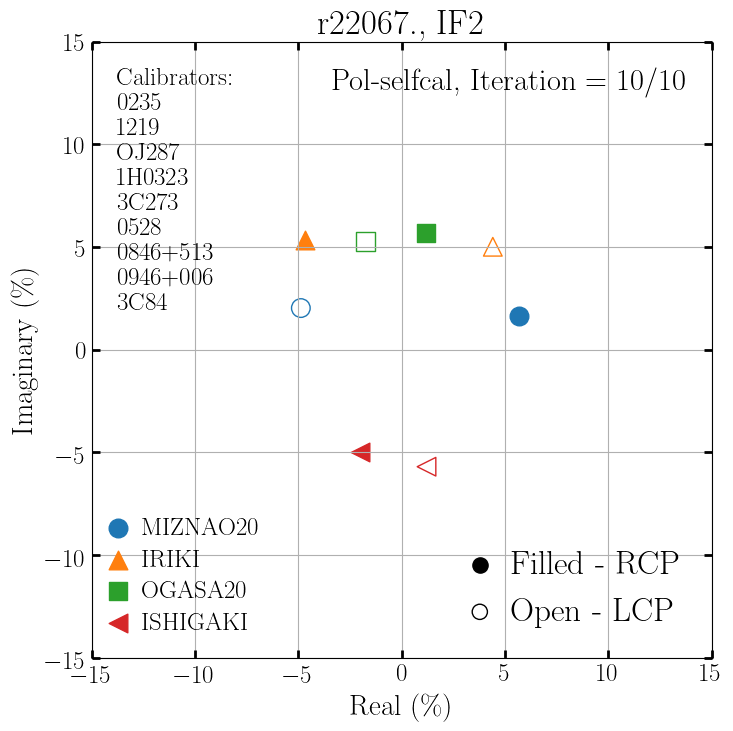}
 \end{center}
  \label{fig:two}
 \end{minipage}
 \hspace{+0.3\columnwidth}
 \begin{minipage}{0.1\hsize}
 \begin{center}
  \includegraphics[width=45mm]{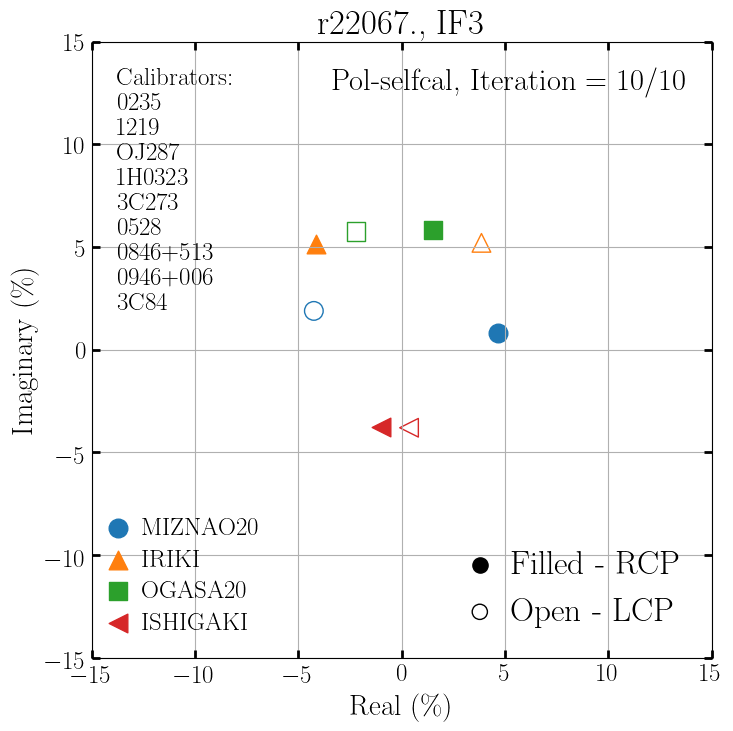}
 \end{center}
  \label{fig:three}
 \end{minipage}
 \hspace{+0.3\columnwidth}
 \begin{minipage}{0.1\hsize}
 \begin{center}
  \includegraphics[width=45mm]{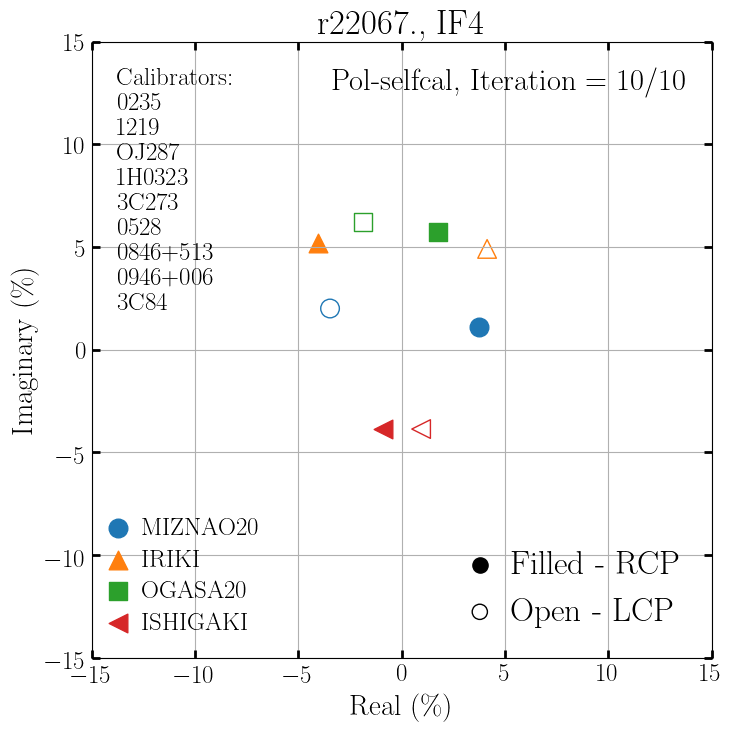}
 \end{center}
  \label{fig:three}
 \end{minipage}
 \caption{K-band D-term results for each of the four 512\,MHz subnands (IF1,IF2, IF3 and IF4) obtained with GPCAL. In each panel, the solutions for different stations are highlighted in different colors, and open/filled circles are RCP/LCP, respectively.}
 \label{fig:gpdtermk}
\end{figure*}

\begin{figure*}[htbp]
 \begin{minipage}{0.1\hsize}
  \begin{center}
   \includegraphics[width=45mm]{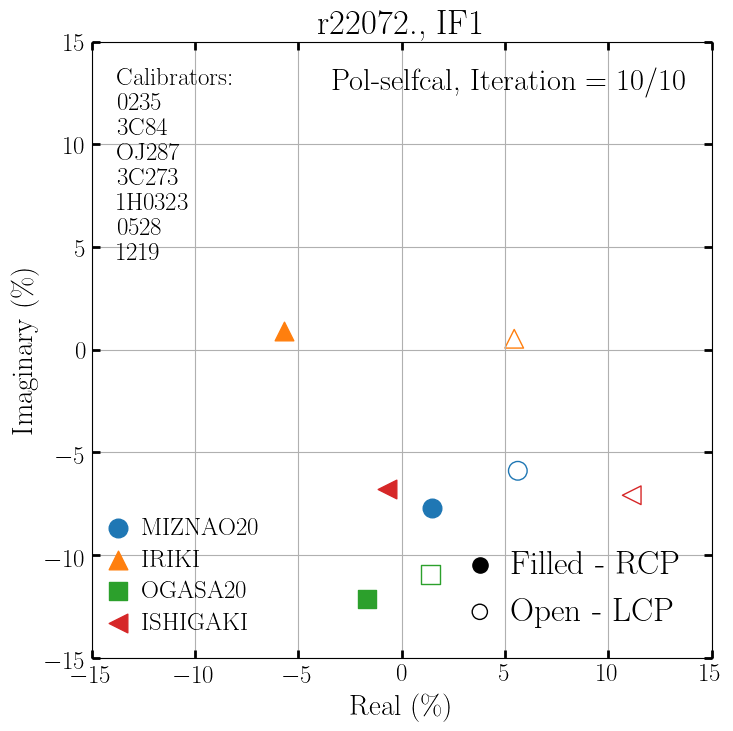}
  \end{center}
  \label{fig:one}
 \end{minipage}
 \hspace{+0.3\columnwidth}
 \begin{minipage}{0.1\hsize}
 \begin{center}
  \includegraphics[width=45mm]{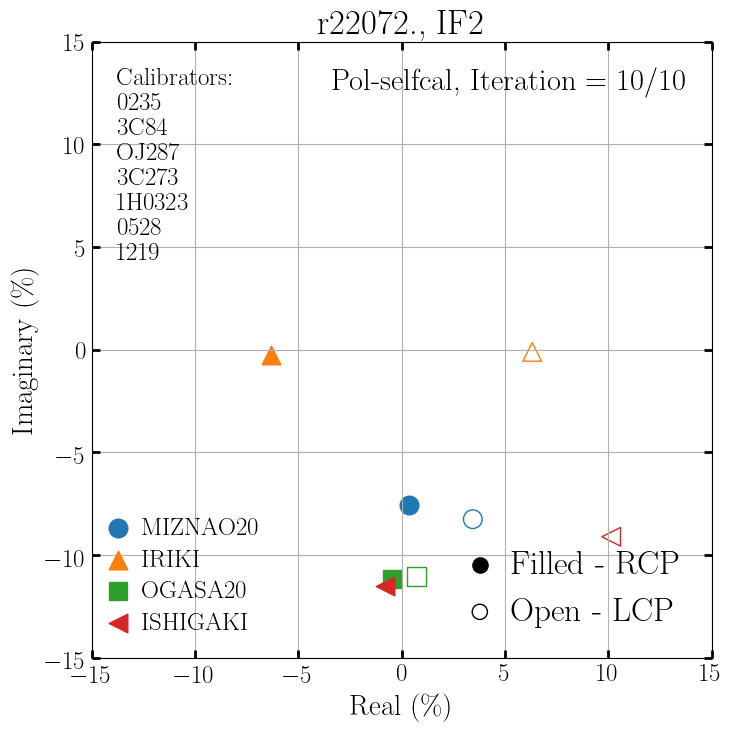}
 \end{center}
  \label{fig:two}
 \end{minipage}
 \hspace{+0.3\columnwidth}
 \begin{minipage}{0.1\hsize}
 \begin{center}
  \includegraphics[width=45mm]{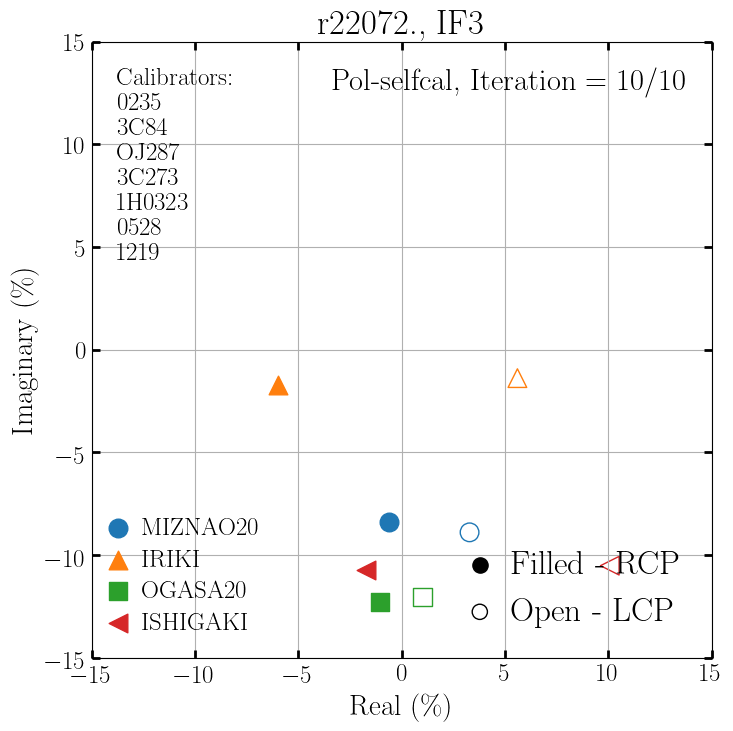}
 \end{center}
  \label{fig:three}
 \end{minipage}
 \hspace{+0.3\columnwidth}
 \begin{minipage}{0.1\hsize}
 \begin{center}
  \includegraphics[width=45mm]{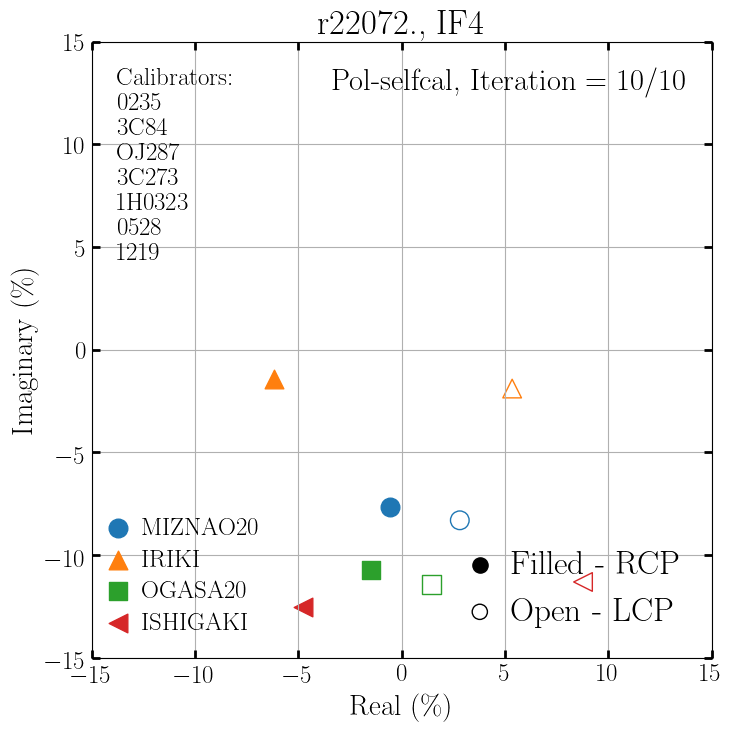}
 \end{center}
  \label{fig:three}
 \end{minipage}
 \caption{Same as Figure\,\ref{fig:gpdtermk} but for Q band.}
 \label{fig:gpdtermq}
\end{figure*}

\begin{acknowledgments}
We thank the referee for his/her valuable comments and suggestions, which signigicantly improved our manuscript. This research was funded by JSPS KAKENHI grant number JP15H03644, JP18KK0090, JP21H04488, JP22H00157, JP21H01137, JP19H01943, and JST SPRING grant number JPMJSP2108. We gratefully acknowledge the support of the Mizusawa VLBI Observatory staff for their assistance with the observing. VERA is operated by National Astronomical Observatory
of Japan. This study makes use of VLBA data from the VLBA-BU Blazar Monitoring Program (BEAM-ME and VLBA-BU-BLAZAR; http://www.bu.edu/blazars/BEAM-ME.html), funded by NASA through the Fermi Guest Investigator Program. The VLBA is an instrument of the National Radio Astronomy Observatory. The National Radio Astronomy Observatory is a facility of the National Science Foundation operated by Associated Universities, Inc.
\end{acknowledgments}

\appendix
\section{D-term Calibration of VERA data} \label{sec:appendix-dterm}
We make use of the Generalized Polarization CALibration pipeline \citep[GPCAL;][]{Park2021} to obtain the D-terms of VERA. This task finds optimal D-term values for each antenna by fitting the polarization model to multiple calibrators simultaneously with a proper consideration of visibility weights among different sources. We observed multiple mildly polarized bright sources over a wide range of parallactic angles so that we can better distinguish the instrumental polarization from the source-intrinsic polarization. In Figure\,\ref{fig:gpdtermk} and \ref{fig:gpdtermq} we show the derived D-terms for each antenna, polarization and subband for epochs R22067A (K band) and R22072B (Q band), respectively. For K/Q bands, 9/7 sources are fit simultaneously (see Figure\,\ref{fig:gpdtermk} and \ref{fig:gpdtermq} for the source list). In K band, the derived D-term solutions are confirmed to be within 4--7\% across the four telescopes, with a tendency that D-term amplitudes overall become smaller at the higher frequency subbands. This indicates that the frequency dependence of D-terms over a wide passband are also properly calibrated in the K-band data. In Q band, the derived D-term amplitudes are somewhat larger than those in K band (especially at Ogasawara and Ishigaki-jima), but still largely constrained to be 5--13\%. 

In \citet{Hagiwara2022}, we also analyzed the same data but using the traditional LPCAL method for a complementary/consistency check. In this analysis we derived D-terms using each 3 calibrator (OJ287, PKS\,0235+164 and 3C84) separately, which allowed us to calculate the standard deviation of derived D-term for each antenna. We obtained the standard deviations to be less than 1--2\%. In GPCAL, we used total 9/7 calibrators for the Dterm validation. Therefore, considering the increase in calibrators, we estimate that the accuracy of D-terms derived from GPCAL is 0.9--1\%.

\section{Polarized images of calibrators}\label{sec:polimage}
\begin{figure*}[htbp]
 \begin{center}
 \includegraphics[width=1.0\linewidth]{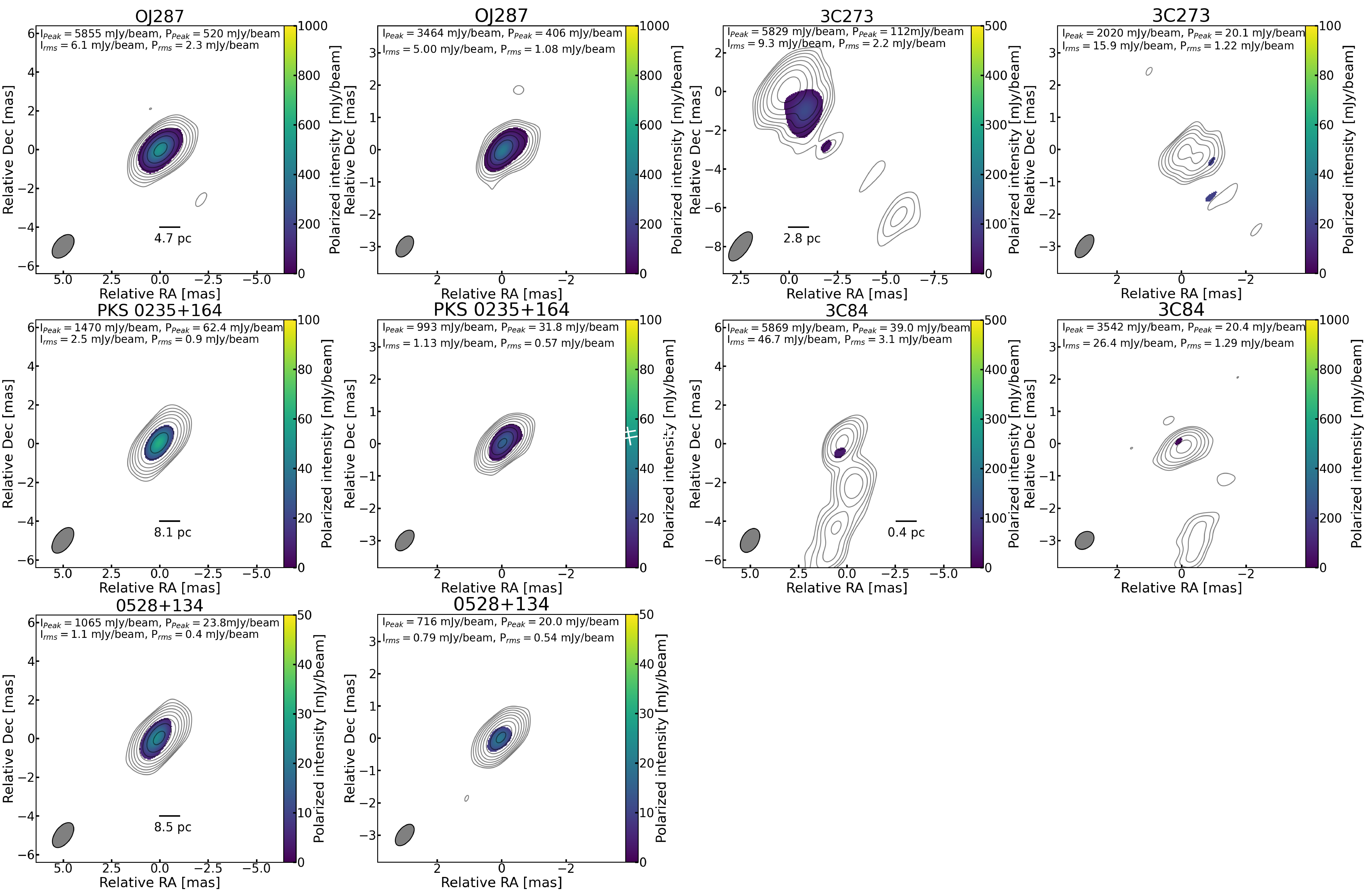}
  \caption{Stokes $I$ and linear-polarization images of calibrators. Contours indicate Stokes $I$ intensity distributions while polarization intensity distributions are color-coded. K band images are shown in the first and third columns, while Q band images are displayed in the second and fourth columns.}
  \label{fig:calimage}
 \end{center}
\end{figure*}

In Figure\,\ref{fig:calimage}, we show linear-polarization images our calibrators obtained with VERA at K and Q bands overlaid on Stokes $I$ intensity contours. Same as Figure\,\ref{fig:images}, here we show the images integrated over the four subbands. Strong polarization signals are seen in the core of OJ\,287, PKS\,0235+164 and 0528+134. For 3C\,273, we detect peak polarization fluxes slightly in the downstream side of of the jet, the characteristics of which is often seen in FSRQ sources. The nucleus of radio galaxy 3C\,84 is well known as a very weakly polarized source at these frequencies. Nevertheless, in both K and Q bands we recover patchy polarization features near the core at $>$10$\sigma$.  
The RM of 3C84 is actively discussed in the literature and RM values of the order of $10^5\,\mathrm{rad\,m^{-2}}$ in the core or extended jet were reported at 43\,GHz and higher frequencies~\citep{Plambeck2014, Nagai2017, Kim2019}. On the other hand, the core RM values obtained in our VERA data at $\leq$43\,GHz ($2.2 \times10^3\,\mathrm{rad\,m^{-2}}$) appear to be smaller than those reported in the literature. The exact reason for the difference is still unclear, but possible RM variability or frequency-dependent nature of RM could be associated with the observed difference. More dedicated observations would be needed to better understand the peculiar polarimetric properties of 3C84.
We detected polarized signal from the core in PKS\,0235+164, OJ287, 0528+134 and 3C84. On the other hand, the 3C273 polarization signal was observed from the jet downstream of the core region.

\section{Evaluations of polarization-related errors}
Here we describe a set of formulas that are used to calculate various error terms associated with a linear-polarization map. Our estimation procedures are based on \citet{Roberts1994} and \citet{Hovatta2012}.
The errors of linear polarization intensity are estimated as
\begin{equation}
    \sigma_{\mathrm{P}} = \frac{\sigma_{Q}+\sigma_{U}}{2}
\end{equation}
\begin{equation}
    \sigma_{EVPA} = \frac{\sigma_{\mathrm{P}}}{2p}
\end{equation}
We ignore the error contribution from the CLEAN procedure because all our observed sources have simple structure. 

\begin{equation}
    \sigma_{Q,U} = (\sigma_{rms}^2 + \sigma_{Dterm}^2)^{1/2},
\end{equation}

\begin{equation}
    \sigma_{\text {Dterm }}=\Delta_{\mathrm{Dterm}}\left(\frac{I^2+\left(0.3 \times I_{\text {peak }}\right)^2}{N_{\text {ant }} \times N_{\text {IF }} \times N_{\text {scan }}}\right)^{1 / 2}
\end{equation}
In our case, $N_{\mathrm{ant}} = 4$, and $N_{\mathrm{IF}}= 1$ when inband RM is obtained. While we set $N_{\mathrm{IF}} = 4$ when we derived RM in Q band. $N_{\mathrm{scan}}$ is depends on each source. 

\section{Intrinsic RM distributions}
\begin{figure*}[htbp]
\centering \includegraphics[scale=0.3]{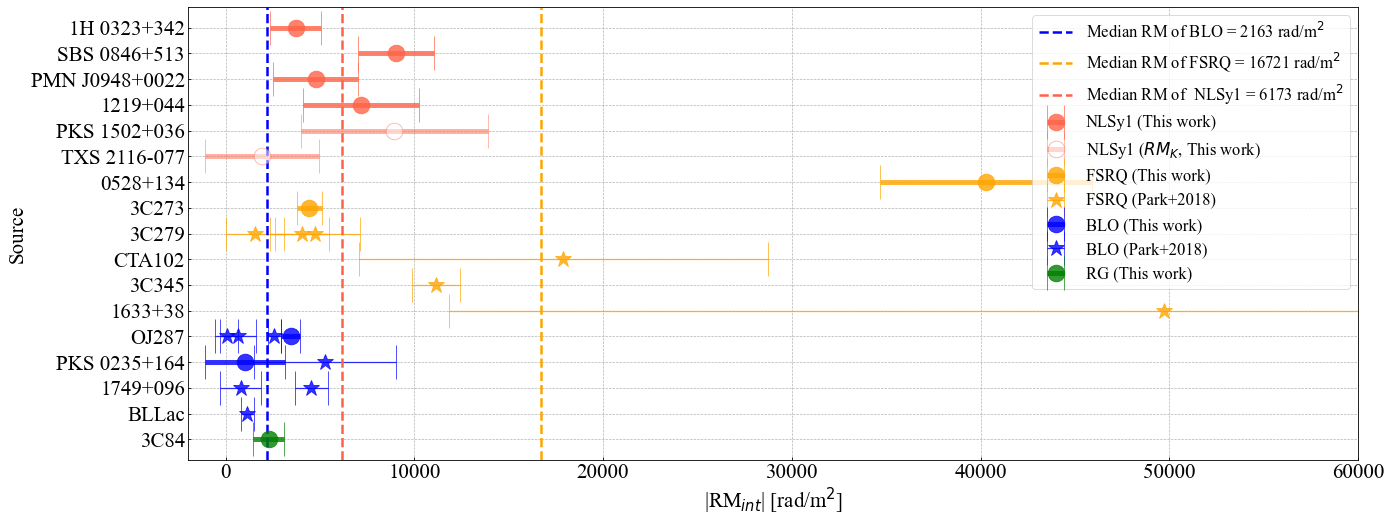}
\caption{Same as Figure\,\ref{fig:RMdistribution} but intrinsic (redshift-corrrected) RM values are plotted instead of the observed ones.}
\label{fig:intRMdist}
\end{figure*}

\begin{figure*}
    \centering
    \includegraphics{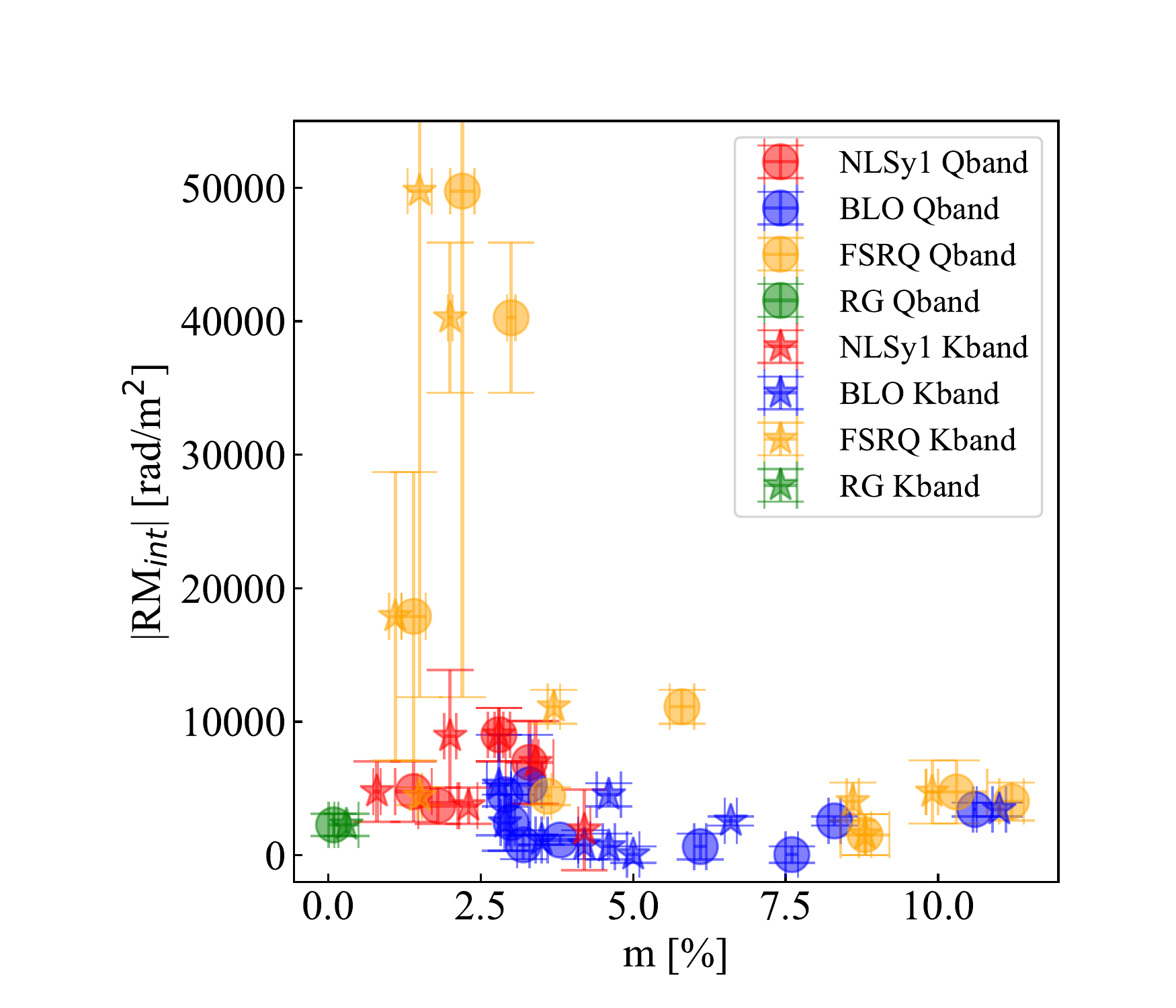}
    \caption{Same as Figure\,\ref{fig:coreRM} but the vertical axis displays intrinsic (redshift-corrected) RM values instead of the observed ones.}
    \label{fig:coreintRM}
\end{figure*}

As an alternative version of Figure\,\ref{fig:RMdistribution}, we create Figure\,\ref{fig:intRMdist}, where intrinsic (redshift-corrected) RM values are compared instead of the observed ones. Similarly, Figure\,\ref{fig:coreintRM} complements Figure\,\ref{fig:coreRM}, where the vertical axis adopts intrinsic RM values instead of the observed ones.

\bibliography{sample631}{}
\bibliographystyle{aasjournal}

\end{document}